\documentclass[a4paper,11pt]{article}
\usepackage{jheppub} % for details on the use of the package, please see the JINST-author-manual
\usepackage{lineno}
\usepackage[T1]{fontenc} % if needed
\usepackage[dvipsnames]{xcolor}
\usepackage{slashed}
\usepackage[normalem]{ulem}
\usepackage{booktabs}
\newsavebox{\tablebox}

\usepackage{orcidlink}

\usepackage{multirow}
\usepackage{comment}
\usepackage{overpic}
\graphicspath{{./figures/}}

\input{belle2-symbols.tex}
\def\DpSCS {\ensuremath{\Dp\to\KS\Km\pip\pip}\xspace}
\def\DsCF {\ensuremath{\Dsp\to\KS\Km\pip\pip}\xspace}
\def\DpDsp     {\ensuremath{\D^+_{(\squark)}}\xspace}
\def\DDsDecay {\ensuremath{\DpDsp\to\KS\Km\pip\pip}\xspace}

\def\ATbar {\ensuremath{\kern 0.2em\overline{\kern -0.2em A}_{T}}\xspace}
\def\ATodd {\ensuremath{{\cal A}_{C\!P}^{\rm TP}}\xspace}

\def\Acp {\ensuremath{{\cal A}_{C\!P}}\xspace}
\def\acpx {\ensuremath{{\cal A}_{C\!P}^{X}}\xspace}

\def\pythsix  {\mbox{\textsc{Pythia6}}\xspace}
\def\pytheight {\mbox{\textsc{Pythia8}}\xspace}

\arxivnumber{2409.15777} % if you have one

\title{\bf {\boldmath Search for $\CP$ violation in $\DDsDecay$ decays using triple and quadruple products}}

\collaboration{The Belle and Belle II Collaborations}
  \author{L.~Aggarwal\,\orcidlink{0000-0002-0909-7537},} % 10083
  \author{H.~Ahmed\,\orcidlink{0000-0003-3976-7498},} % 11323
  \author{H.~Aihara\,\orcidlink{0000-0002-1907-5964},} % 2223
  \author{N.~Akopov\,\orcidlink{0000-0002-4425-2096},} % 9443
  \author{A.~Aloisio\,\orcidlink{0000-0002-3883-6693},} % 2194
  \author{N.~Althubiti\,\orcidlink{0000-0003-1513-0409},} % 21524
  \author{N.~Anh~Ky\,\orcidlink{0000-0003-0471-197X},} % 2218
  \author{D.~M.~Asner\,\orcidlink{0000-0002-1586-5790},} % 4684
  \author{H.~Atmacan\,\orcidlink{0000-0003-2435-501X},} % 2538
  \author{V.~Aushev\,\orcidlink{0000-0002-8588-5308},} % 2155
% ...
\abstract{
We perform the first search for $\CP$ violation in 
${D_{(s)}^{+}\to K_{S}^{0}K^{-}\pi^{+}\pi^{+}}$ decays. 
We use a combined data set from the Belle and Belle II 
experiments, which study $e^+e^-$ collisions at 
center-of-mass energies at or near
the $\Upsilon(4S)$ resonance. We use 980~fb$^{-1}$ of 
data from Belle and 428~fb$^{-1}$ of data from Belle~II.
We measure six $\CP$-violating asymmetries that are based on 
triple products and quadruple products of the momenta of 
final-state particles, 
and also the particles' helicity angles.
We obtain a precision at the level
of 0.5\% for $\Dp\to\KS\Km\pip\pip$ decays, and better than 
0.3\% for $\Dsp\to\KS\Km\pip\pip$ decays. 
No evidence of $\CP$ violation is found. 
Our results for the triple-product asymmetries are the
most precise to date for singly-Cabibbo-suppressed $D^+$ decays.
Our results for the other asymmetries 
are the first such measurements performed for charm decays.
}

\begin{document}
\preprint{
\vbox{ 
%\hbox{\href{https://docs.belle2.org/record/4366}{BELLE2-PUB-DRAFT-2024-011} v6.2}
%\hbox{\href{https://docs.belle2.org/record/3825}{BELLE2-NOTE-PH-2023-049}}
%\hbox{Authors: L.~K.~Li, A.~J.~Schwartz, K.~Kinoshita}
%\hbox{RC: M.~Staric, M.~Nayak, X.~B.~Ji}
%\hbox{PC reader: Y. Sakai}
\vskip-20pt
\hbox{Belle II Preprint 2024-025}
\hbox{KEK Preprint 2024-24}
% UCHEP-24-05 
}
}

\maketitle
\flushbottom

\section{\boldmath Introduction}
The violation of charge-conjugation plus parity~($\CP$) symmetry 
holds significant importance in particle physics, as it is essential 
for explaining the matter-antimatter imbalance in the Universe~\cite{Sakharov:1967dj}. 
Within the Standard Model~(SM), $\CP$ violation~($\CPV$) arises from a 
complex phase in the Cabibbo-Kobayashi-Maskawa (CKM) 
matrix~\cite{Kobayashi:1973fv}. 
However, this source is insufficient to explain the observed matter-antimatter 
imbalance, and thus we conclude there must be other sources of $\CPV$, 
presumably arising from new physics~(NP) beyond the SM. 

In the SM, $\CPV$ in the charm sector arises at the level 
of $\mathcal{O}(10^{-3})$ or less~\cite{Brod:2011re,Cheng:2019ggx,Cheng:2021yrn},
and observing $\CPV$ significantly above this level would be interpreted as a 
sign of~NP~\cite{Dery:2019ysp,Delepine:2019cpp,Chala:2019fdb,Saur:2020rgd,Bause:2022jes,Bediaga:2022sxw,Lenz:2023rlq,Pich:2023kim,Iguro:2024uuw}. 
Various methods have been used to search for $\CPV$ in charm
decays~\cite{HFLAV:2024ctg}:
measuring differences in decay widths~\cite{LHCb:2021rdn,Belle:2021dfa,Belle:2021ygw,LHCb:2022pxf,BESIII:2022qrs,LHCb:2022lry}, 
searching for differences in decay-time distributions or regions of phase-space~\cite{LHCb:2024jpt,LHCb:2020zkk,LHCb:2023qne}, 
measuring triple-product asymmetries~\cite{LHCb:2014djq,Belle:2023qio,Belle:2023bzn,Belle:2023str}, 
and measuring decay asymmetry parameters in baryonic decays~\cite{CLEO:2004txf,Belle:2021crz,Belle:2022uod}, 
and performing amplitude analyses
and an ``energy test''~\cite{Williams:2011cd,LHCb:2016qbq,LHCb:2023mwc}
of multi-body decays.
To date, the only observation of $\CPV$ in charm
was reported by the LHCb experiment~\cite{LHCb:2019hro}, which 
measured a difference between the asymmetries $A^{KK}_{\CP}$ and 
$A^{\pi\pi}_{\CP}$ for $D^0\to K^+K^-$ and $D^0\to\pip\pim$ decays. The 
asymmetry $A_{\CP}$ is the difference between partial widths of a $D$ and a $\overline{D}$ divided by their sum;
a nonzero value mainly arises from {\it direct\/} $\CPV$, i.e., 
interference between two or more decay amplitudes to
the same final state. The effect is proportional to 
$\sin\delta\sin\phi$, where $\delta$ is the strong phase difference
between the two amplitudes, and $\phi$ is the weak phase difference.

Four-body decays of charmed mesons typically proceed via 
intermediate resonances, and the corresponding amplitudes
interfere with one another. Thus, these decays
offer a promising opportunity to observe $\CPV$.
One observable sensitive to $\CPV$ in  
$D\to h_1\,h_2\,h_3\,h_4$ ($h\!=\!\pi,K,\eta$, etc.)
decays is the ``triple product''
$C_{\rm TP} \equiv (\vec{p}_1 \times \vec{p}_2)\cdot \vec{p}_3$,
where the momenta of the final state particles
$\vec{p}_{1,2,3}$ are measured in the rest frame of the~$D$.
A difference between the 
$C_{\rm TP}$ distribution for a $D$ decay and
the corresponding $\overline{C}_{\rm TP}$ distribution for a $\overline{D}$ decay
is $\CP$-violating. To quantify the difference,
the asymmetry about zero for the $C_{\rm TP}$ 
distribution is compared to the asymmetry for 
the $\overline{C}_{\rm TP}$ distribution; 
the difference is denoted $\ATodd$.
Unlike the asymmetry in partial widths $A_{\CP}$, 
the triple-product asymmetry
$\ATodd$ is proportional to 
$\cos\delta\sin\phi$~\cite{Valencia:1988it,Datta:2003mj,Wang:2022fih},
i.e., it reaches its maximum value at $\delta\!=\!0$. 
The observables $C_{\rm TP}$ and $\overline{C}_{\rm TP}$ 
are $T$-odd, and thus nonzero values could indicate $T$ violation.

\begin{figure}[!hbtp]
  \begin{center}% 
  \begin{overpic}[width=0.32\textwidth]{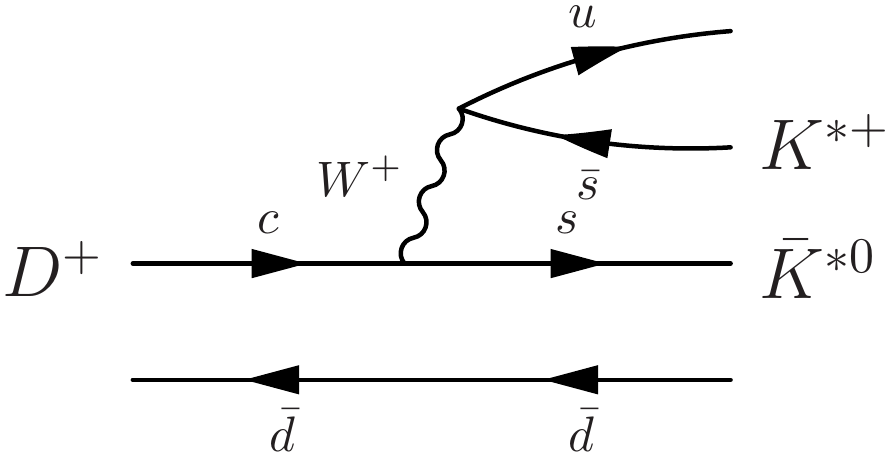}%
  \end{overpic}~~
  \begin{overpic}[width=0.32\textwidth]{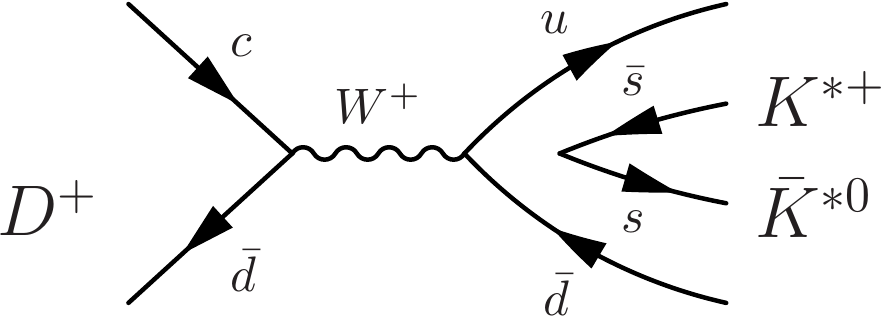}%
  \end{overpic}~~
  \begin{overpic}[width=0.32\textwidth]{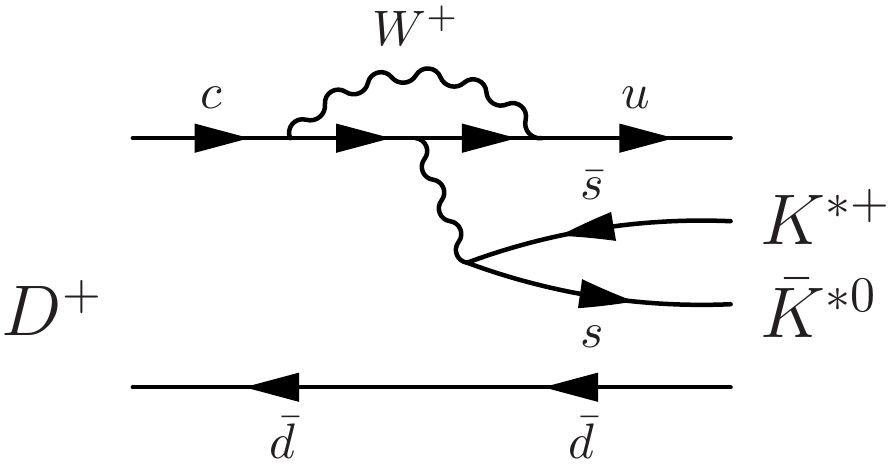}%
  \end{overpic}%  
  \caption{Tree (left), annihilation (center), and ``penguin'' (right) 
  diagrams contributing to $D^+\to \overline{K}{}^{*0}K^{*+}$. This decay
  is expected to be the dominant process for the four-body decay $\DpSCS$.
  \label{fig:FeymKstKst}}
  \end{center}
\end{figure}

Several experiments have searched for $\CPV$ using 
triple-product asymmetries in four-body $D$ decays~\cite{HFLAV:2022esi}, 
thus far without success. Here we measure the triple-product asymmetry 
for the four-body decays\footnote{For the event selections and analysis procedure, charge-conjugate modes are implicitly included unless noted otherwise.}
$\DpSCS$ and $\DsCF$. The first mode is singly 
Cabibbo-suppressed, like the $D^0\to K^+K^-$ and $D^0\to\pi^+\pi^-$ decays for which $\CPV$ was observed.
The second mode ($D_s^+$ decay) is Cabibbo-favored; in this case the SM prediction 
for \CPV\ is negligible, and observing \CPV\ would indicate new physics.
 The dominant intermediate process, $D^+\to\Kstarzb\Kstarp$, 
involves tree, annihilation, and ``penguin'' amplitudes as 
shown in figure~\ref{fig:FeymKstKst}. These amplitudes interfere 
with one another, potentially giving rise to $\CPV$. In addition 
to measuring the triple-product asymmetry, we also measure 
the asymmetry for the ``quadruple product'' 
$C_{\rm QP} \equiv (\vec{p}_1 \times \vec{p}_2)\cdot (\vec{p}_3\times \vec{p}_4)$.
Unlike the triple product, $C_{\rm QP}$ is $T$-even and also $P$-even.
Quadruple-product asymmetries are discussed as a 
method for measuring $\CPV$ in refs.~\cite{Durieux:2016nqr,Wang:2022fih,Durieux:2015zwa}.
Finally, we measure asymmetries in helicity angle
distributions, which can also exhibit $\CP$ violation~\cite{Durieux:2015zwa,Zhang:2022iye,Wang:2024qff}.
To date, quadruple-product asymmetries and asymmetries in helicity angle
distributions have not yet been studied for multibody charm decays.

\section{\boldmath Methodology}
\begin{figure}[!hbt]
  \begin{center}% 
  \begin{overpic}[width=0.8\textwidth]{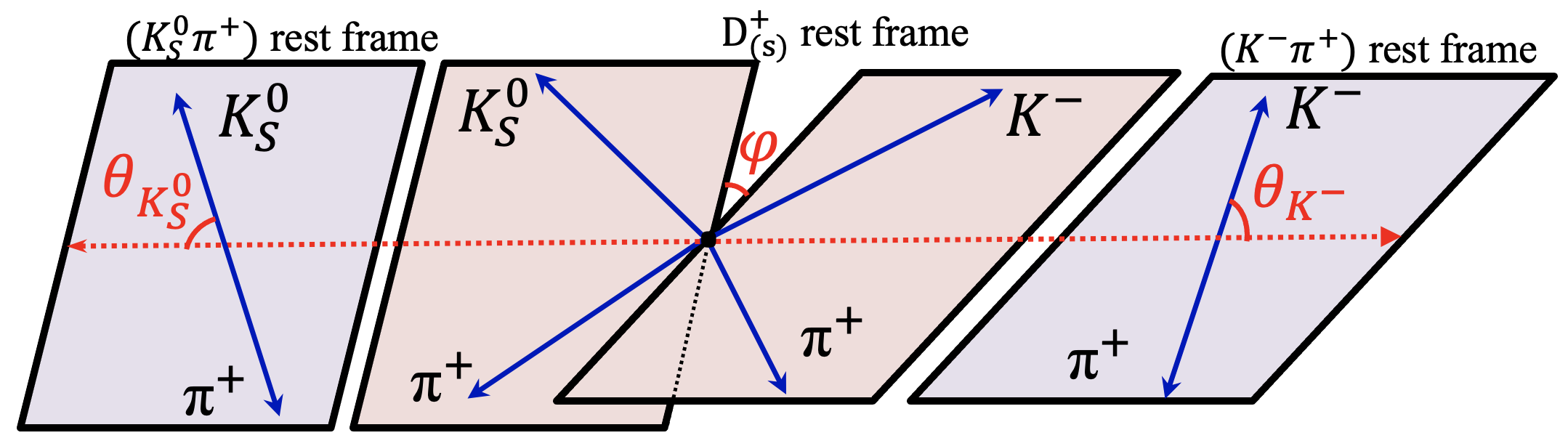}%
  \end{overpic}%
  \caption{\label{fig:CTHelicity}
   Decay topology for $D^+_{(s)}\to\KS\Km\pip\pip$. The innermost decay planes,
   as drawn, correspond to the $D^+_{(s)}$ rest frame; the outermost decay planes, 
   as drawn, correspond to the $K_S^0\,\pi^+$ and $K^-\pi^+$ rest frames.
The red dashed line denotes the direction of the $(K\pi)$
system's momentum in the $D^+_{(s)}$ rest frame, and the direction 
opposite the $D^+_{(s)}$ momentum in the $(K\pi)$ system's rest frame.}
  \end{center}
\end{figure}

The topology of $\DDsDecay$ decays is shown in figure~\ref{fig:CTHelicity}.
The angle in the $D^+_{(s)}$ rest frame between the decay planes of the $\KS\,\pip$ 
pair and the $\Km\pip$ pair is denoted~$\varphi$. Also shown are two
helicity angles, $\theta_{\KS}$ and $\theta_{K^-}$. These are defined as
the angle in the $\KS\,\pip$ or $K^-\pi^+$ rest frame between the kaon 
momentum and the direction opposite that of the $D^+_{(s)}$ momentum.

The triple and quadruple products are defined as:
\begin{eqnarray}
C_{\rm TP} & = & (\vec{p}_{K^-} \times \vec{p}_{\pi^+_h})\cdot \vec{p}_{\KS}\,, \\
C_{\rm QP} & = & (\vec{p}_{K^-} \times \vec{p}_{\pi^+_h})\cdot  (\vec{p}_{\KS}\times \vec{p}_{\pi^+_l})\,,
\end{eqnarray}
where the subscripts ``$h$'' or ``$l$'' denote
the pion with the higher or lower momentum, respectively. 
All momenta are measured in the $D_{(s)}^+$ rest 
frame.\footnote{For $C_{\rm TP}$, the choice of three out of four final state momenta does not impact the measurement, as the total momentum sums to zero. For $C_{\rm QP}$, the combinations are chosen to be similar to $C_{\rm TP}$.}
In addition to searching for $\CPV$ using $C_{\rm TP}$ and $C_{\rm QP}$, 
we also use the product $C_{\rm TP}C_{\rm QP}$ and three functions of 
the helicity angles: 
$\cos\theta_{\KS}\cos\theta_{\Km}$, 
$C_{\rm TP}\cos\theta_{\KS}\cos\theta_{\Km}$, and 
$C_{\rm QP}\cos\theta_{\KS}\cos\theta_{\Km}$. The 
signs of the observables $C_{\rm QP}$, $C_{\rm TP}$, and $C_{\rm TP}C_{\rm QP}$ 
are the same as those of
$\cos\varphi$, $\sin\varphi$, and $\sin(2\varphi)$, respectively. 
The sign of $C_{\rm TP}\cos\theta_{\KS}\cos\theta_{\Km}$ 
is the same as that of
an interference term in the angular distribution of $D\to V_a V_b$, $V\to P_1 P_2$
decays~\cite{Durieux:2015zwa}: $d_{1,0}^2(\theta_a) d_{1,0}^2(\theta_b)\sin\varphi$, 
which is proportional to $\sin(2\theta_a)\sin(2\theta_b)\sin\varphi$. 
The term $C_{\rm QP}\cos\theta_{\KS}\cos\theta_{\Km}$ has the same sign as another 
interference term, $d_{1,0}^2(\theta_a) d_{1,0}^2(\theta_b)\cos\varphi$, which is proportional to 
$\sin(2\theta_a)\sin(2\theta_b)\cos\varphi$. 
In these expressions, $V$ and $P$ denote vector and pseudoscalar mesons, and 
the $d$'s denote Wigner $d$ functions~\cite{Jacob:1959at,Chung:1997jn}.
Thus, a \CP\ asymmetry in $C_{\rm TP}\cos\theta_{\KS}\cos\theta_{\Km}$ 
or $C_{\rm QP}\cos\theta_{\KS}\cos\theta_{\Km}$ indicates an asymmetry in the
corresponding interference terms. As interference terms are sensitive to new 
physics amplitudes, such an asymmetry could indicate new physics.
We thus measure six observables,
$X=C_{\rm TP}$, $C_{\rm QP}$, ${C_{\rm TP}C_{\rm QP}}$, 
$\cos\theta_{\KS}\cos\theta_{\Km}$, $C_{\rm TP}\cos\theta_{\KS}\cos\theta_{\Km}$, 
and $C_{\rm QP}\cos\theta_{\KS}\cos\theta_{\Km}$, the signs of which correspond 
to the signs of different combinations of $\sin\varphi$, $\cos\varphi$, 
$\cos\theta_{\KS}$, and $\cos\theta_{\Km}$. 

For each observable, we measure the asymmetry about zero
for both $D^+_{(s)}$ and $D^-_{(s)}$ decays:
\begin{eqnarray}
A_X(D^+_{(s)}) & \equiv & \frac{ N(D_{(s)}^+,\,X>0) - N(D_{(s)}^+,\,X<0) }{ N(D_{(s)}^+,\,X>0) + N(D_{(s)}^+,\,X<0) } \label{eqn:Ax} \\
\nonumber \\
A_{\overline{X}}(D^-_{(s)}) & \equiv & 
\frac{ N(D_{(s)}^-,\,\overline{X}>0) - N(D_{(s)}^-,\,\overline{X}<0) }
{ N(D_{(s)}^-,\,\overline{X}>0) + N(D_{(s)}^-,\,\overline{X}<0)} \,,   \label{eqn:Axb} 
\end{eqnarray} 
where $N$ denotes the yield of $D_{(s)}^+$ or $D_{(s)}^-$ decays. 
For the observables proportional to $C_{\rm TP}$, 
which is the product of a polar vector and an axial vector,
$\overline{X}$ is multiplied by $-1$
to account for the oddness of $C_{\rm TP}$ under a parity transformation.
With this extra factor,
$A_X$ and $A_{\overline{X}}$ are 
$\CP$-conjugate quantities for all the observables, and any 
difference violates~$\CP$. We quantify the difference with
\begin{eqnarray}
\acpx & \equiv & 
\frac{A_X(D^+_{(s)}) - A_{\overline{X}}(D^-_{(s)})}{2}\,,
\label{eqn:acpx}
\end{eqnarray} 
where $\acpx\neq 0$ indicates $\CP$ violation.

\section{Detector and data set}
\label{sec:data}

The Belle detector~\cite{bib:BelleDetector} was a large-solid-angle spectrometer 
that operated at the KEKB asymmetric-energy $e^+e^-$ collider~\cite{bib:KEKB,bib:KEKB2}.
It had a cylindrical geometry and consisted of a silicon vertex detector (SVD), 
a central drift chamber~(CDC), 
an array of aerogel threshold Cherenkov counters~(ACC), a barrel-like arrangement 
of time-of-flight scintillation counters~(TOF), and an electromagnetic calorimeter~(ECL) 
based on CsI(Tl) crystals. These components were located inside a superconducting 
solenoid coil that provided a 1.5~T magnetic field. The iron flux-return of the magnet 
was instrumented with resistive-plate chambers~(KLM) to detect muons and $K^0_L$ mesons.
More details on the Belle detector are provided in refs.~\cite{bib:BelleDetector,bib:BelleDetector2}.

The Belle II detector~\cite{Abe:2010gxa}, operates at the SuperKEKB asymmetric-energy 
$e^+e^-$ collider~\cite{Akai:2018mbz} and also has a cylindrical geometry. It includes 
a two-layer silicon-pixel detector~(PXD)~\cite{Belle-IIDEPFETPXD:2019cub,Belle-IIDEPFET:2021pib} 
surrounded by a four-layer double-sided 
silicon-strip detector~\cite{Belle-IISVD:2022upf} and a 56-layer central drift chamber. 
Only one sixth of the second layer of the PXD was installed for the data analyzed here. 
Outside the CDC is a time-of-propagation counter~(TOP)~\cite{Kotchetkov:2018qzw} in 
the central region and an aerogel-based ring-imaging Cherenkov counter~(ARICH) 
in the forward region.  
Surrounding the TOP and ARICH is the ECL 
and the 1.5~T superconducting solenoid magnet previously used in Belle. 
The magnet's flux return is instrumented with resistive-plate 
chambers and plastic scintillator modules to detect muons, 
$K^0_L$ mesons, and neutrons. More details on the Belle~II 
detector are provided in ref.~\cite{Abe:2010gxa}.
The symmetry axis of both Belle and Belle~II detectors is
defined as the $z$ axis, and in both cases it is almost coincident with the electron beam direction.
Both magnetic fields run parallel to the $z$ axis. 

This analysis uses both Belle and Belle~II data sets,
980~$\invfb$ recorded by Belle~\cite{bib:BelleDetector2} and 428~$\invfb$ recorded by Belle~II~\cite{Belle-II:2024vuc}.
The majority of the data (73\% for Belle, 85\% for Belle~II) were
recorded at an $e^+e^-$ center-of-mass (c.m.) energy corresponding
to the $\Upsilon(4S)$ resonance. The remaining data were recorded
at energies slightly above and below the $\Upsilon(4S)$ resonance 
and at other $\Upsilon(nS)$ ($n=1,2,3,5$) resonances.
The analysis is performed using the 
Belle~II analysis software framework~(BASF2)~\cite{Kuhr:2018lps},
with Belle data converted to BASF2 format with the B2BII
software package~\cite{Gelb:2018agf}.
We use Monte Carlo~(MC) simulated events to optimize 
selection criteria, study backgrounds, and calculate 
reconstruction efficiencies. The MC samples are generated using
\evtgen~\cite{Lange:2001uf}, and the detector response is 
simulated using \geantb~\cite{Lange:2001uf} for Belle and
\geant~\cite{GEANT4:2002zbu} for Belle~II. The continuum 
process $\epem\to\qqbar$, where $q=u,\,d,\,s,\,c$, is 
generated using 
\pythsix~\cite{Sjostrand:2006za} 
for Belle and \pytheight~\cite{Sjostrand:2014zea} 
and \kkmc~\cite{Jadach:1999vf} for Belle~II. Final-state radiation of 
charged particles is simulated with \photos~\cite{Barberio:1990ms}.

\section{Event selection}
\label{sec:selection}
Event selection criteria are chosen to maximize a figure-of-merit 
${\rm FOM} = N_S/\sqrt{N_S + N_B}$, where
$N_S$ is the signal yield expected in a region 
$|M(D)-m_D|<13~{\rm MeV}/c^2$, and  
$N_B$ is the background yield expected in this region.
Here, $M(D)=M(\KS K^-\pi^+\pi^+)$ is the invariant mass of 
the $D_{(s)}^+$ candidate, $m_D$ is the nominal $D_{(s)}^+$ mass~\cite{bib:PDG2024}, 
and the range corresponds to about $2.5\sigma$ in resolution.
These yields are obtained from MC simulation; the background yield from MC is scaled by the ratio 
of yields between data and MC in the sideband $20~{\rm MeV}/c^2 < |M(D) - m_D| < 40~{\rm MeV}/c^2$ 
to obtain $N_B$.

We require that signal candidates pass the following selection criteria:
charged tracks must lie within the CDC acceptance (an angular coverage of 
$17^{\circ}\leq\theta\leq150^{\circ}$), have at least one CDC hit, 
and have a distance-of-closest-approach to the $e^+e^-$ 
interaction point (IP) of less than 2.0~cm along the $z$ direction
and of less than 0.5~cm in the $x$-$y$ (transverse) plane. 
For each track, we calculate particle identification 
likelihoods ($\mathcal{L}$) for $\pi$ and $K$
particle hypotheses
using information from the CDC, ACC, and TOF detectors in Belle, and 
mainly from the CDC, TOP, and ARICH detectors in Belle~II. 
Tracks satisfying a likelihood ratio 
$\mathcal{L}_{\pi}/(\mathcal{L}_{\pi}+\mathcal{L}_K)\!>\!0.6$ 
are identified as pions, and tracks satisfying a ratio
$\mathcal{L}_K/(\mathcal{L}_K+\mathcal{L}_{\pi})\!>\!0.6$ 
are identified as kaons. 
Tracks that have high electron or muon likelihoods
are rejected, where these lepton likelihoods
are determined using information mainly from the ECL and KLM detectors, 
respectively~\cite{Hanagaki:2001fz,Abashian:2002bd,Milesi:2020esq}. 
All these requirements have an average efficiency of 
91\% for kaons and 95\% for pions at Belle, and 
87\% for kaons and 90\% for pions at Belle~II.

Candidate $\KS$ mesons are reconstructed from pairs of oppositely-charged tracks assumed to be
pions. To suppress non-$\KS$ background at Belle, 
a neural network~(NN)~\cite{Feindt:2006pm} is employed.
This NN uses 13 input variables~\cite{Belle:2018xst}; the most discriminating are 
the $\KS$ momentum in the laboratory frame, 
the distance to the IP in the $x$-$y$ plane for each track, and
the $\KS$ flight length in the $x$-$y$ plane.
To suppress non-$\KS$ background at Belle~II, the pion candidates are required to originate from 
a common vertex with a fit quality $\chi^2_{\KS}<100$. 
The invariant mass of the $\KS\to\pip\pim$ candidate is required 
to lie within $10~{\rm MeV}/c^2$ of the nominal $\KS$ mass~\cite{bib:PDG2024}.
These requirements retain about 97\% (Belle)
and 92\% (Belle~II) of $K_S^0$ decays while rejecting 
more than 98\% of non-$\KS$ background such as $D^+\to\Km\pip\pim\pip\pim$.
The difference in efficiency is partly due to a slightly longer tail in the 
$M(\pi^+\pi^-)$ resolution function for Belle~II data.

We reconstruct $D_{(s)}^+$ candidates by combining a $\KS$ candidate, a $K^-$ track, 
and two $\pi^+$ tracks. A vertex fitting algorithm~\cite{Krohn:2019dlq} 
is applied to the entire decay chain, subject to a mass constraint for the $\KS$
and a constraint that the $D_{(s)}^+$ momentum originate from the IP.
%%%It requires daughters to originate from a common vertex (referred to as $D_{(s)}^+$ decay vertex 
%%%and $\KS$ product vertex) and constraints $D_{(s)}^+$ being produced at IP, along with a 
The resulting goodness-of-fit $\chi^2$ is required to be less than 30 for $D^+$ decays
and less than 40 for $D_s^+$ decays. 
The $D_{(s)}^+$ flight significance, defined as $L_{D}/\sigma^{}_{L^{}_D}$,
is required to be greater than 3.0 (0.5) for $D^+$ ($D^+_s$) decays at Belle, 
and greater than 4.5 (2.0) for $D^+$ ($D^+_s$) decays at Belle~II.
Although the Belle~II criteria are tighter, both sets of criteria
have similar efficiencies
as a consequence of the superior vertex resolution of the 
Belle~II PXD detector and significantly higher background rejection.

After the $D_{(s)}^+$ reconstruction,
the $\KS$ flight length ($L$) is calculated as the projection 
of the displacement vector joining the $D_{(s)}^+$ and $\KS$ decay vertices 
onto the direction of the $\KS$ momentum. The uncertainty $\sigma^{}_L$ is 
calculated by propagating the uncertainties on the vertex positions and 
the $\KS$ momentum, accounting for their correlations. The $\KS$ flight 
significance $L/\sigma^{}_L$ is required to be greater than a minimum
value, which is optimized separately for Belle and Belle~II (greater 
than 10 at Belle and greater than 20 at Belle~II).

We define a scaled momentum for the $D_{(s)}^+$ candidate as 
$x_p\!\equiv p^{}_{D}/p^{}_{\rm max}$, where
$p^{}_D$ is the $D$ momentum and 
$p^{}_{\rm max} = \sqrt{E^2_{\rm beam} - M(D)^2 c^4}/c$.
Here, $E^{}_{\rm beam}$ is the beam energy, and 
both $E^{}_{\rm beam}$ and $p^{}_{D}$ are evaluated in the $e^+e^-$ center-of-mass frame. 
To suppress large backgrounds originating from the continuum and from $B$ 
decays, we require $x_p>0.5$. 
We also require that the invariant mass of the $D_{(s)}^+$ candidate satisfy
$\bigl[ M(D)\!- m^{}_D\bigr]\!\in\!(-50, 40)$~MeV/$c^2$. 
For $D_s^+$ candidates, an additional background arising
from $D^{*+} \to D^0 (\to\KS\Km\pip)\pip$ is present in 
the upper $M(\KS K^-\pi^+\pi^+)$ sideband. To suppress this background, 
we require that, for each $\pi^+$ candidate, 
$\bigl[ M(\KS\Km\pip\pip)-M(\KS\Km\pip)\bigr] >151.57$~MeV/$c^2$.
This requirement eliminates essentially all such background while
preserving more than 99\% of signal $D_s^+$
decays. The overall reconstruction efficiencies at Belle
are 6\% for $D^+$ decays and 9\% for $D^+_s$ decays;
the efficiencies at Belle~II are very similar: 
7\% for $D^+$ decays and 9\% for $D^+_s$ decays.

\section{$\ensuremath{{\cal A}_{C\!P}^{X}}\xspace$ measurement}

We measure the asymmetries $\ensuremath{{\cal A}_{C\!P}^{X}}$,
separately for Belle and Belle~II,
in two steps. We first perform a fit to the $M(D)$ distributions of 
a combined sample of $(D^+ + D^-)$ decays, and 
a combined sample of $(D^+_{s} + D^-_{s})$ decays.
From this fit we determine the shape of the $M(D)$ probability 
density function (PDF) for signal decays.
We subsequently perform a second fit for the 
asymmetries~$\acpx$, where the PDF shape for signal decays
is fixed to that obtained from the first fit.

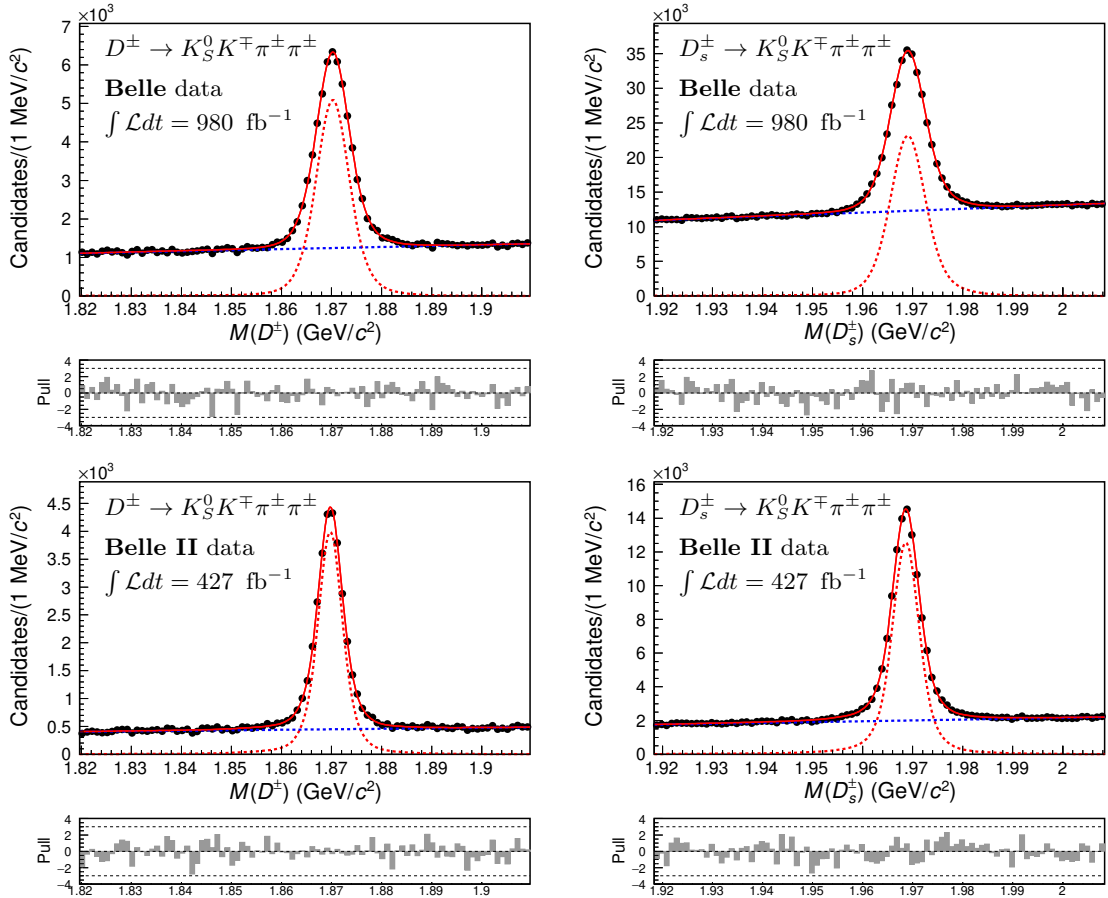
\begin{figure}[!hbtp]
  \begin{center}%   
  \begin{overpic}[width=0.5\textwidth]{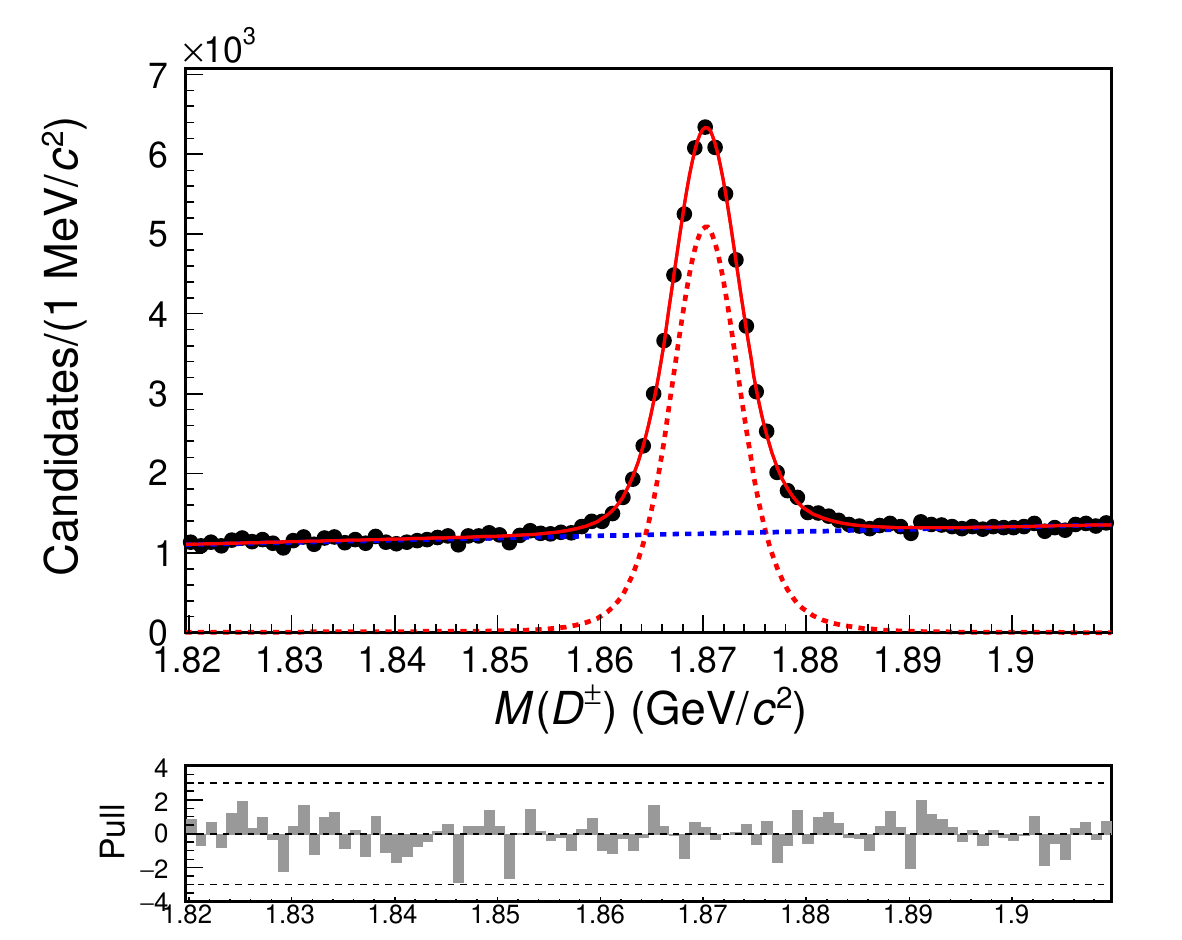}%
  \put(20,68){\footnotesize{$D^\pm\to\KS{}K^\mp\pi^\pm\pi^\pm$}}
  \put(20,61){\footnotesize{{\bf Belle} data}}
  \put(20,55){\footnotesize{$\int\mathcal{L}dt=980~\invfb$}}
  \end{overpic}%
  \begin{overpic}[width=0.5\textwidth]{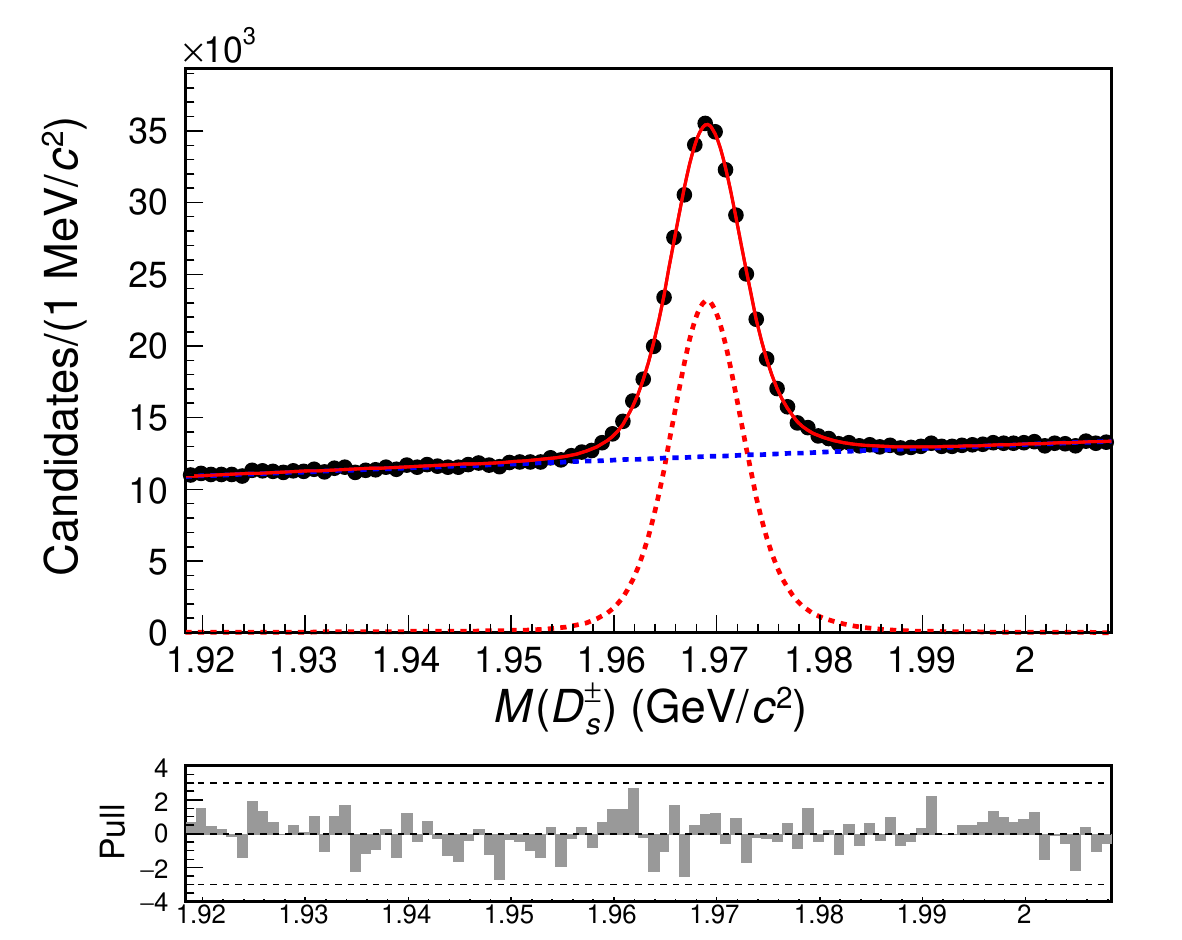}%
  \put(20,68){\footnotesize{$D_s^\pm\to\KS{}K^\mp\pi^\pm\pi^\pm$}}  
  \put(20,61){\footnotesize{{\bf Belle} data}}
  \put(20,55){\footnotesize{$\int\mathcal{L}dt=980~\invfb$}}   
  \end{overpic}\\
  \vskip2pt
  \begin{overpic}[width=0.5\textwidth]{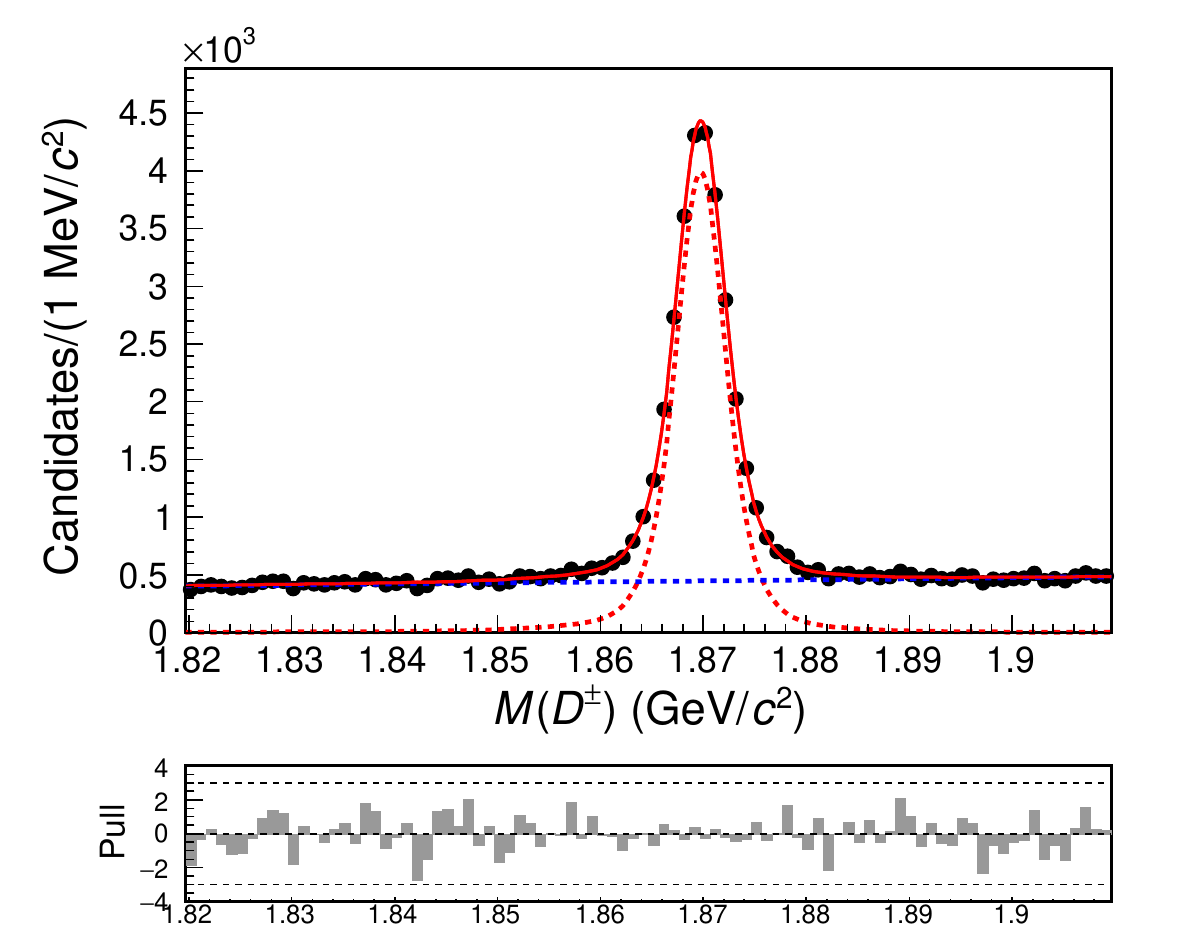}%
  \put(20,68){\footnotesize{$D^\pm\to\KS{}K^\mp\pi^\pm\pi^\pm$}}
  \put(20,61){\footnotesize{{\bf Belle~II} data}}
  \put(20,55){\footnotesize{$\int\mathcal{L}dt=427~\invfb$}} 
  \end{overpic}%
  \begin{overpic}[width=0.5\textwidth]{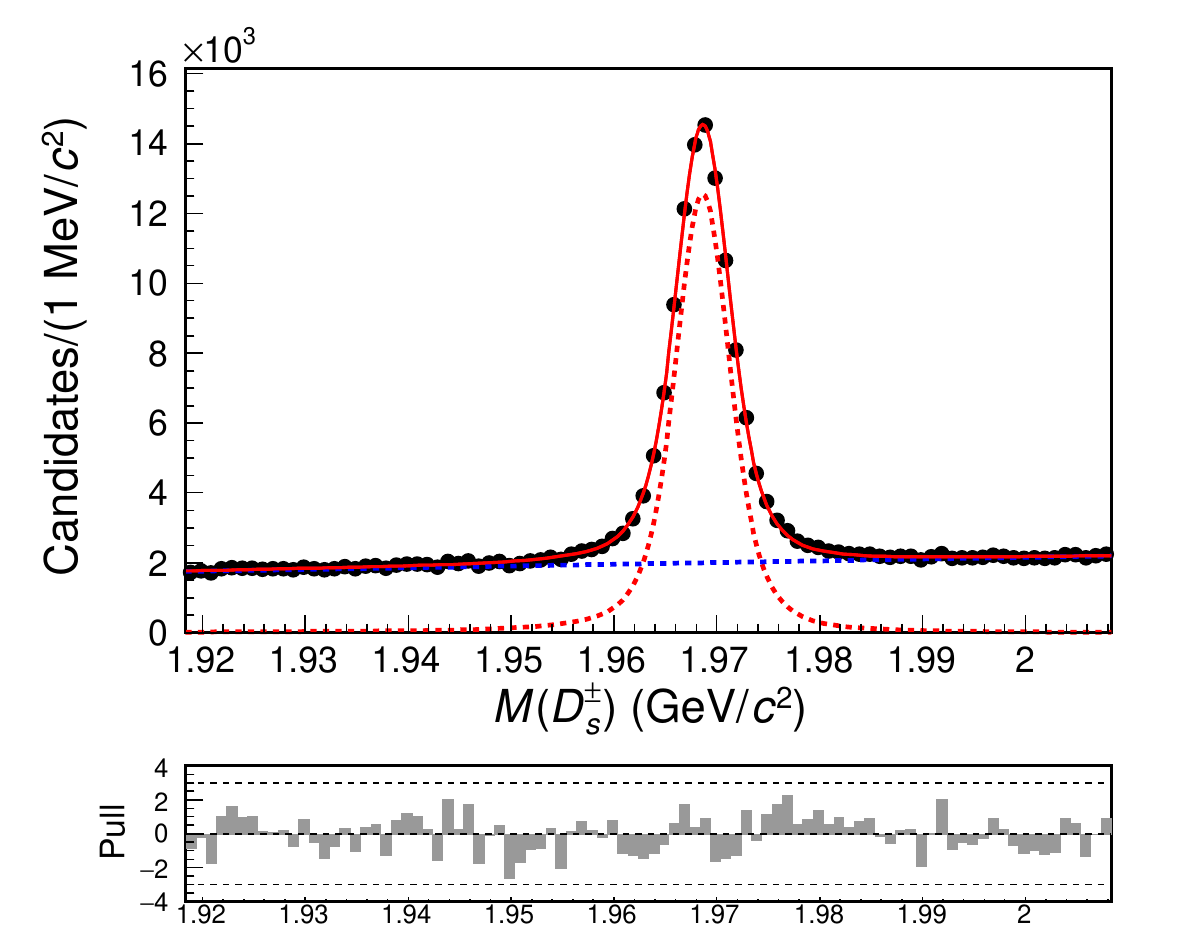}%
  \put(20,68){\footnotesize{$D_s^\pm\to\KS{}K^\mp\pi^\pm\pi^\pm$}}  
  \put(20,61){\footnotesize{{\bf Belle~II} data}}
  \put(20,55){\footnotesize{$\int\mathcal{L}dt=427~\invfb$}}
  \end{overpic}
  \caption{\label{fig:finalYields}
     Invariant mass distributions for 
     $D_{(s)}^\pm\to\KS K^\mp\pi^\pm\pi^\pm$
     candidates (points with error bars),
     with fit results overlaid. The red dashed curves show the fitted signal, 
     and the blue dash-dotted curves show the fitted background. The top plots show Belle data, and
     the bottom plots show Belle~II data. 
     The smaller panels show the pull distributions.} 
  \end{center}
\end{figure}

The PDF for signal is taken to be the sum of a double Gaussian 
and two (for $D^+$) or three (for $D_s^+$) asymmetric Gaussians.
These functions share a common mean parameter but have distinct 
width parameters. 
All parameters, including 
the relative fractions of the Gaussians,
are taken from MC simulation.
However,
a shift $\delta_\mu$ to the mean and a common scaling factor
$k_\sigma$ for the widths are floated in the first fit
(separately for $D^+$ and $D_s^+$)
to account for small differences between data and MC simulation.
A possible difference between the $M(D)$ shapes 
for $D^+_{(s)}$ and $D^-_{(s)}$ is considered when evaluating 
systematic uncertainties.
For background, the PDF is taken to be a straight 
line, which over the range of interest describes the data well.
The fitted $M(D)$ distributions and projections of 
the first fit result 
are shown in figure~\ref{fig:finalYields}.
Also shown are the pull distributions,
where the pull is defined as
$({\rm data\ yield}\!-\!{\rm fitted\ yield})/({\rm {data\ uncertainty}})$.
The resulting signal and background yields
in a window $\pm 10$~MeV/$c^2$ 
around the nominal $D_{(s)}^+$ mass~\cite{bib:PDG2024}
are listed in table~\ref{tab:FinalYields}. The ratio of signal to 
background for Belle~II is higher than that for Belle due to the 
improved resolution on the $D^+_{(s)}$ mass and 
flight length~$L^{}_D$.

\begin{table}[!htbp]
\begin{center}
\caption{\label{tab:FinalYields} Fitted signal and background yields in a window $\pm 10$~MeV/$c^2$ around the nominal 
$D_{(s)}^+$ mass~\cite{bib:PDG2024}. 
The uncertainties listed are statistical.}
\vskip0.10in
\begin{lrbox}{\tablebox}
\renewcommand{\arraystretch}{1.1}
\begin{tabular}{l|cc|cc} \toprule  	
\multirow{2}{*}{Component}	
    & \multicolumn{2}{c|}{$D^\pm\to\KS K^\mp\pi^\pm\pi^\pm$}	
	& \multicolumn{2}{c}{$D^\pm_s\to\KS K^\mp\pi^\pm\pi^\pm$}	
	\\ \cline{2-5}
        &  Belle 	&    Belle~II	&   Belle 	& Belle~II	\\ \midrule  
%Signal  & $~\,46073\pm 301$   & $28224\pm 209$	& $223182\pm 826$	& $~\,98214\pm 419$   \\ %\cline{2-8}
%Background   & $110555\pm 394$	& $39937\pm 236$	& $1091491\pm 1245$	& $177743\pm 505$  \\ \midrule
%Ratio ($N_{\rm sig}/N_{\rm bkg}$)  &    0.42 & 0.71 & 0.20 & 0.55 \\ \midrule
% \textcolor{blue}{signal in SR ($N^{\rm SR}_{\rm sig}$)}	& $44048\pm 288$	& $26738\pm 199$		& $210743\pm 780$		& $92000\pm 393$		\\ %\cline{2-8}
% \textcolor{blue}{background in SR ($N^{\rm SR}_{\rm bkg}$)}	& $24844\pm 88~$	& $8964\pm 53$		& $245285\pm 280$	& $39997\pm 114$		\\ 
%\textcolor{blue}{Ratio ($N^{\rm SR}_{\rm sig}/N^{\rm SR}_{\rm bkg}$)} 		& 	1.8		&	3.0					&	0.9				& 2.3			\\	
Signal ($N^{}_{\rm sig}$)	& $44048\pm 288$	& $26738\pm 199$		& $210743\pm 780$		& $92000\pm 393$		\\ %\cline{2-8}
Background ($N^{}_{\rm bkg}$)	& $24844\pm 88~$	& $8964\pm 53$		& $245285\pm 280$	& $39997\pm 114$		\\ 
Ratio ($N^{}_{\rm sig}/N^{}_{\rm bkg}$) 		& 	1.8		&	3.0		&	0.9				& 2.3			\\	
\bottomrule
\end{tabular}
  \end{lrbox}
  \scalebox{0.95}{\usebox{\tablebox}}
\end{center}  
\end{table} 

For the second fit to determine $\acpx$, 
the data is divided into four subsamples as determined by 
the charge of $D_{(s)}^{\pm}$ and the sign of~$X$. For all 
subsamples, the background slope and calibration factors 
$\delta_\mu$ and $k_\sigma$ are fixed to the values obtained 
from the first fit.
%\section{$\ensuremath{{\cal A}_{C\!P}^{X}}\xspace$ measurement}
%We measure the $\CP$-violating asymmetries $\acpx$ in eq.~(\ref{eqn:acpx}) by performing 
%an unbinned maximum likelihood fit to the invariant mass $M(D)$ distributions.
The yields of the subsamples are expressed as:
\begin{eqnarray}
N(D_{(s)}^+, X>0) & = & \frac{N_{+}}{2} (1 + A_X) \\
N(D_{(s)}^+, X<0) & = & \frac{N_{+}}{2} (1 - A_X) \\
N(D_{(s)}^-, \overline{X}>0) & = & \frac{N_{-}}{2} (1 + A_X - 2\acpx ) \\
N(D_{(s)}^-, \overline{X}<0) & = & \frac{N_{-}}{2} (1 - A_X + 2\acpx ) \,.
\end{eqnarray} 
Here, $N_+ = N(D_{(s)}^+,\,X>0) + N(D_{(s)}^+,\,X<0)$
and $N_{-} = N(D_{(s)}^-,\,\overline{X}>0) + 
N(D_{(s)}^-,\,\overline{X}<0)$ represent the total
signal yields for $D_{(s)}^+$ and $D_{(s)}^-$, respectively;
$A_X$ denotes the asymmetry for~$X$ 
as defined in eq.~\ref{eqn:Ax}; and 
$\acpx$ denotes the $\CP$-violating parameter 
as defined in eq.~\ref{eqn:acpx}. 
We perform a simultaneous fit to the $M(D)$ distributions of
these four subsamples to extract parameters 
$N_+$, $N_{-}$, $A_X$, and $\acpx$. 
We test the fitting procedure on MC samples generated with different
input values of $\acpx$; in all cases we obtain fitted values for 
$\acpx$ consistent with the input values.

The results for $\acpx$ are listed in table~\ref{tab:FinalAcp} along with 
systematic uncertainties. 
As the signs of our observables are uncorrelated, all measured asymmetries are statistically independent. 
The systematic uncertainties listed, also uncorrelated, are discussed in the next section.
The fitted $M(D)$ distributions are shown 
in appendix~\ref{app:MDfit}.
The Belle~II distributions for $C_{\rm TP}$ and $-\overline{C}_{\rm TP}$, 
and for $C_{\rm QP}$ and $\overline{C}_{\rm QP}$, 
are shown in appendix~\ref{app:Xplots}; the 
corresponding distributions for Belle look very similar. No difference 
in shapes is seen between $D^+_{(s)}$ and $D^-_{(s)}$ decays.

\begin{table}[!htbp]
\begin{center}
\caption{\label{tab:FinalAcp} 
Results for $\acpx$ 
in units of $10^{-3}$ 
for $D_{(s)}^{+}\to\KS\Km\pip\pip$ decays.
The first and second uncertainties listed 
are statistical and systematic, respectively.
The last column lists the significance of the combined 
result from $\acpx = 0$.}
\vskip0.10in
\begin{lrbox}{\tablebox}
\begin{tabular}{ll|cccc} \toprule  
&$X$ & $\acpx$ Belle  &  $\acpx$ Belle~II &   
Combined $\acpx$ & Significance		\\ 
%& & ($10^{-3}$) & ($10^{-3}$) & ($10^{-3}$) & \\
\midrule  
\multirow{6}{*}{$D^+$}
&$C_{\rm TP}$
& $-4.0\pm 5.9\pm 3.0$		& $-0.2\pm 7.0\pm 1.8$		& $-2.3\pm 4.5\pm 1.5$	 & $0.5\sigma$		\\  
&$C_{\rm QP}$
& $-1.0\pm 5.9\pm 2.5$		& $-0.4\pm 7.0\pm 2.4$		& $-0.7\pm 4.5\pm 1.7$	 & $0.2\sigma$		\\  
&$C_{\rm TP}\,C_{\rm QP}$
& $+6.4\pm 5.9\pm 2.2$		& $+0.6\pm 7.0\pm 1.3$		& $+3.9\pm 4.5\pm 1.1$	 & $0.8\sigma$		\\  
&$\cos\theta_{\KS}\cos\theta_{\Km}$
& $-4.7\pm 5.9\pm 3.0$		& $-0.6\pm 6.9\pm 3.0$		& $-2.9\pm 4.5\pm 2.1$	 & $0.6\sigma$		\\  
&$C_{\rm TP}\cos\theta_{\KS}\cos\theta_{\Km}$ 
& $+1.9\pm 5.9\pm 2.0$		& $-0.2\pm 7.0\pm 1.9$		& $+1.0\pm 4.5\pm 1.4$	 & $0.2\sigma$		\\  
&$C_{\rm QP}\cos\theta_{\KS}\cos\theta_{\Km}$
& $+14.9\pm 5.9\pm 1.4$		& $+7.0\pm 7.0\pm 1.6$		& $+11.6\pm 4.5\pm 1.1$	 & $2.5\sigma$		\\  \midrule 
\multirow{6}{*}{$D_s^+$}
&$C_{\rm TP}$	& $-0.3\pm 3.1\pm 1.3$		& $+1.0\pm 3.9\pm 1.1$		& $+0.2\pm 2.4\pm 0.8$	  & $0.1\sigma$		\\ 
&$C_{\rm QP}$		& $+0.6\pm 3.1\pm 1.2$		& $+2.0\pm 3.9\pm 1.4$		& $+1.1\pm 2.4\pm 0.9$	  & $0.4\sigma$		\\ 
&$C_{\rm TP}\,C_{\rm QP}$		& $+1.5\pm 3.2\pm 1.4$		& $-2.7\pm 3.9\pm 1.7$		& $-0.2\pm 2.5\pm 1.1$	  & $0.1\sigma$		\\ 
&$\cos\theta_{\KS}\cos\theta_{\Km}$	& $-3.7\pm 3.1\pm 1.1$		& $-6.3\pm 3.9\pm 1.2$		& $-4.7\pm 2.4\pm 0.8$	  & $1.8\sigma$		\\ 
&$C_{\rm TP}\cos\theta_{\KS}\cos\theta_{\Km}$ 	& $-4.4\pm 3.2\pm 1.4$		& $+0.8\pm 3.9\pm 1.4$		& $-2.2\pm 2.5\pm 1.0$	  & $0.8\sigma$		\\ 
&$C_{\rm QP}\cos\theta_{\KS}\cos\theta_{\Km}$	& $-1.6\pm 3.1\pm 1.3$		& $-0.0\pm 3.9\pm 1.7$		& $-1.0\pm 2.4\pm 1.0$	  & $0.4\sigma$		\\ \bottomrule
\end{tabular}
  \end{lrbox}
  \scalebox{0.88}{\usebox{\tablebox}}
\end{center}  
\end{table}

\section{Systematic uncertainties}
\label{sec:systAcp}

Many systematic uncertainties for the measurements of $X$ and $\overline{X}$ cancel 
in the ratios $A_{X}$ (eq.~\ref{eqn:Ax}) and $A_{\overline{X}}$ (eq.~\ref{eqn:Axb}),
or in the difference between them (eq.~\ref{eqn:acpx}). 
The uncertainties that do not cancel are listed in table~\ref{tab:AcpSys} and discussed 
below.

\begin{table}[!htbp]
\begin{center}
\caption{\label{tab:AcpSys}
Systematic uncertainties (absolute) for $\acpx$ in units of $10^{-3}$ 
in $D_{(s)}^{+}\to\KS\Km\pip\pip$ decays, where
$X\!= C_{\rm TP}~(1)$, 
$C_{\rm QP}~(2)$, 
$C_{\rm TP}C_{\rm QP}~(3)$,
$\cos\theta_{\KS}\cos\theta_{\Km}~(4)$, 
$C_{\rm TP}\cos\theta_{\KS}\cos\theta_{\Km}~(5)$, and 
$C_{\rm QP}\cos\theta_{\KS}\cos\theta_{\Km}~(6)$.}
\vskip0.10in
\begin{lrbox}{\tablebox}
\begin{tabular}{l|cccccc|cccccc} \toprule  
 \multirow{2}{*}{Source} 
	& \multicolumn{6}{c|}{$\DpSCS$ at Belle}	& \multicolumn{6}{c}{$\DpSCS$ at Belle~II} \\ \cline{2-13}
	& ~(1) & ~(2) & ~(3) & ~(4) & ~(5) & ~(6) 	& ~(1) & ~(2) & ~(3) & ~(4) & ~(5) & ~(6)  \\ \midrule
$X$-dependent efficiency 	
		&  3.0 & 2.4 & 1.9 & 2.8 & 1.8 & 1.4		& 1.2 & 2.4 & 1.1 & 2.6 & 1.5 & 1.3  \\		
$X$-resolution asymmetry 
		&  0.2 & 0.7 & 0.4  & 0.7  & 0.6 & 0.3 	& 0.7 & 0.1 & 0.1  & 0.9  & 0.2 & 0.7 	\\
Signal/background PDF
		&  0.0 & 0.0 & 0.0  & 0.0  & 0.0 & 0.0 	& 0.0 & 0.0 & 0.0  & 0.0  & 0.0 & 0.0 	\\
Simultaneous fit bias	
		& 0.2 & 0.2 & 0.1 & 0.2 & 0.1 & 0.2 		& 0.2 & 0.2 & 0.2 & 0.1 & 0.2 & 0.2	\\
$D_s^+$ feeddown background
		& 0.4 & 0.3 & 1.0 & 0.7 &   0.1 & 0.2 		& 1.1 & 0.4 & 0.6 & 1.1 & 1.2 & 0.6	\\ \hline 
Total $\sigma_{\rm syst}$	
		&  3.0 & 2.5 & 2.2 & 3.0 & 2.0 & 1.4		& 1.8 & 2.4 & 1.3 & 3.0 & 1.9 & 1.6  	 \\	\midrule   
\multirow{2}{*}{Source} 
	& \multicolumn{6}{c|}{$\DsCF$ at Belle}		& \multicolumn{6}{c}{$\DsCF$ at Belle~II} \\ \cline{2-13}
	& ~(1) & ~(2) & ~(3) & ~(4) & ~(5) & ~(6) & ~(1)~ & ~(2)~ & ~(3)~ & ~(4)~ & ~(5)~ & ~(6)~ \\ \midrule
$X$-dependent efficiency 	
		& 1.2 & 1.1 & 1.4 & 1.1 & 1.2 & 1.3		& 1.1 & 1.4 & 1.7 & 1.2 & 1.4 & 1.6  	  \\
$X$-resolution asymmetry 
		&  0.6 & 0.5 & 0.1 & 0.2  & 0.8 & 0.3 		& 0.2 & 0.1 & 0.2 & 0.0 & 0.2 & 0.4 	\\
Signal/background PDF
		&  0.0 & 0.0 & 0.0 & 0.0 & 0.0 & 0.0 		& 0.0 & 0.0 & 0.0 & 0.0 & 0.0 & 0.0    \\
Simultaneous fit bias	
		& 0.1 & 0.0 & 0.0 & 0.1 & 0.1 & 0.0		&  0.1 & 0.2 & 0.2 & 0.3 & 0.2 & 0.2	\\ \hline  
Total $\sigma_{\rm syst}$	
		& 1.3 & 1.2 & 1.4 & 1.1 & 1.4 & 1.3		& 1.1 & 1.4 & 1.7 & 1.2 & 1.4 & 1.7  	  \\  \bottomrule
\end{tabular} 
  \end{lrbox}
  \scalebox{0.83}{\usebox{\tablebox}}
\end{center}  
\end{table}  

A difference in 
efficiencies between positive and negative $X$ values
could lead to false asymmetries $A_{X}$ and $A_{\overline{X}}$. 
Such a difference is evaluated using MC simulation by comparing 
reconstructed $X$ distributions with those that were generated. 
For this study, MC signal decays were generated in two ways: according to phase space and with several intermediate resonances.
For $X\!=\!C_{\rm TP}$ and $C_{\rm TP}C_{\rm QP}$, no significant 
$X$-dependence is observed.
For the other observables, 
a small dependence is observed, but it
is the same for $D^+_{(s)}$ and $D^-_{(s)}$ decays 
and thus the effect upon $\acpx$ is small.
To account for this quantitatively, 
we take as a systematic uncertainty
the difference between generated and reconstructed values of $\acpx$.
We include in this uncertainty 
the statistical uncertainty
of the MC samples used.
Finally, we include in this uncertainty any effect upon $\acpx$
arising from a possible difference between 
MC simulation and data for $X$-dependent efficiencies.
We evaluate this as follows.
The asymmetry in detection efficiencies between $K^+$ and $K^-$ 
(denoted $A^K_\varepsilon$) is measured in data 
as a function of momentum and polar angle using
$D^0\to\Km\pip$ and $D_s^+\to\phi\pip$ decays~\cite{Belle:2012ygx},
and the asymmetry in detection efficiencies between $\pi^+$ and 
$\pi^-$ (denoted $A^\pi_\varepsilon$) is measured using 
$D^+\to\Km\pip\pip$ and $D^0\to\Km\pip\piz$ decays~\cite{Belle:2012ygt}.
We scale MC signal events with
weighting factors based on these asymmetries and repeat the fits for $\acpx$. 
For $D^+_{(s)}\to K_S^0 K^-\pi^+\pi^+$, the weighting factor is 
$(1- A_{\eff}^{K})(1+ A_{\eff}^{\pi})(1+ A_{\eff}^{\pi})$; 
for $D^-_{(s)}\to K_S^0 K^+\pi^-\pi^-$, the weighting factor is 
$(1+ A_{\eff}^{K})(1- A_{\eff}^{\pi})(1- A_{\eff}^{\pi})$.
After fitting these reweighted events, the resulting 
change in $\acpx$ is included in the systematic uncertainty for 
$X$-dependent efficiency. This contribution is very small ($<5\!\times\!10^{-4}$).

We consider the effect of
the resolution in $X$. 
Such resolution leads to events migrating from $+X$ values to $-X$ values and vice versa, diluting the measure asymmetry. 
From MC simulation, we find that the net fraction of signal decays changing sign in $X$ is less than 0.1\%.
This corresponds to a dilution factor of $1\!-\!2\!\times\!0.1\%\!=\!0.998$, which is a negligible effect given the statistical uncertainty. 
We also study the effect of a difference in $X$ resolution between positive and negative values. 
From MC simulation, we find that the resolution in~$X$ depends on $X$ but is
well-described by a parabola symmetric about $X\!=\!0$.
%We consider the effect of a difference in $X$ {\it resolution\/}
%between positive and negative values. If such a difference existed, 
%then a different number of events could shift from $-X$ to $+X$ values than from $+X$ to $-X$, 
%corrupting the $A_X$ measurement.
%From MC simulation, we find that the net fraction of 
%signal decays changing sign in $X$ is very small, less than 0.1\%. 
%To evaluate the effect quantitatively, 
%we study the $X$ resolution for different ranges of $X$.
%We find the resolution in $X$ to depend on $X$ but be well-described 
%by a parabola symmetric about $X\!=\!0$. 
The statistical error on this 
parabola is used to generate MC samples with asymmetric resolution; these samples 
are then fitted and the resulting change in $\acpx$ is taken as the 
systematic uncertainty due to a possible asymmetry in $X$ resolution.
%in various slices of the region $\pm10\sigma_{X=0}$ in resolution. 
%These resolutions are well-described by a parabolic function, with the 
%fitted factor of first-order term consistent with zero, indicating good symmetry.
%We further assess the event flip asymmetry between positive-to-negative and 
%negative-to-positive regions of $X$-observables based on signal MC sample.

The uncertainty due to fixed PDF shape parameters is evaluated by varying these
parameters by their uncertainties and repeating the fit. We sample all parameters
simultaneously from Gaussian distributions having widths equal to their uncertainties,
and repeat this sampling and subsequent fitting 1000 times. During this sampling, correlations among the sampled parameters are accounted for.
The 1000 fitted values for $\acpx$ are recorded, and the 
standard deviation of this distribution is taken as the systematic uncertainty 
due to signal and background PDF shapes. These 
uncertainties are very small,
less than $2\times10^{-5}$.

We have studied a possible difference between the $M(D)$ shapes for $D^+_{(s)}$ 
and $D^-_{(s)}$ by separately floating the calibration factors $\delta^{}_\mu$ 
and $k^{}_\sigma$. We find the shifts $\delta^{}_\mu$ to be essentially identical, 
and the scaling factors $k^{}_\sigma$ to be consistent within $1\sigma$, i.e., for 
Belle~II data we obtain $k^{}_{\sigma}(D^+)=1.021\pm 0.013$ and $k^{}_{\sigma}(D^-)=1.027\pm 0.013$.
The changes in $A_{CP}^X$ when using separate calibration factors for $D^+_{(s)}$ and 
$D^-_{(s)}$ are found to be negligible.
We have also studied a possible difference in $M(D)$ shapes for $X\!>\!0$ and 
$X\!<\!0$ and found any such difference negligible.

To assess possible bias in our fitting procedure, we fit a large 
sample of ``toy'' MC events generated by sampling the PDFs used to fit 
the data. The number of generated events for $\pm X$ and $\pm\overline{X}$ 
vary: they are calculated for different input values of $\acpx$, which range 
from $-0.015$ to $0.015$. For each input value, an ensemble of MC events 
is generated. These ensembles are fitted, and the mean value of the 
$\acpx$ results is calculated. Plotting these mean values against the 
input values shows a linear dependence. Fitting these points to a line 
gives a slope consistent with unity and an intercept consistent with zero. 
We assign the difference between our measured value and the 
corresponding input value
as given by the fitted line as a systematic uncertainty due to possible 
fit bias. The uncertainty on the difference is included in this calculation.

For the $D^+\to\KS\Km\pip\pip$ decay mode, we consider the effect of possible background 
from $D_s^+\to\KS\Km\pip\pip\piz$, with the $\pi^0$ missed. This five-body decay is 
unmeasured, and thus to evaluate the effect of this
background, we include a component for it in our fit. We take the shape of its PDF from 
MC simulation and float its yield. The difference between the resulting value of $\acpx$
and our nominal result is taken as a systematic uncertainty. 

Combining all systematic uncertainties in quadrature gives the total systematic
uncertainties listed in table~\ref{tab:AcpSys}. These total uncertainties are also
included in table~\ref{tab:FinalAcp}. For all $\acpx$ measurements, the total 
systematic uncertainties are notably less than the
corresponding statistical uncertainties.

\section{Combined result and summary} 

Table~\ref{tab:FinalAcp} lists our results for the six asymmetries $\acpx$, for both 
Belle and Belle~II. 
As the systematic uncertainties for Belle and Belle~II, as evaluated,
are uncorrelated,
we combine the results from the two data sets using the formulae
\begin{eqnarray}
\acpx & = &    \frac{\Acp^{B1}/\sigma_{B1}^2 + \Acp^{B2}/\sigma_{B2}^2}{1/\sigma_{B1}^2 + 1/\sigma_{B2}^2}
\label{eqn:combined} \\ \nonumber \\
\sigma_{\acpx}  & = &   1/\sqrt{1/\sigma_{B1}^2 + 1/\sigma_{B2}^2}\,,		 
\label{eqn:combinedErr}
\end{eqnarray}
where $\sigma^{}_{B_1}$ and $\sigma^{}_{B_2}$ represent the total uncertainties 
(sum in quadrature of statistical and systematic uncertainties) for Belle 
and Belle~II measurements, respectively.
The combined results are also listed in table~\ref{tab:FinalAcp}; 
the resulting uncertainties are 
at the level of 0.5\% for $D^+$ decays and better
than 0.3\% for $D_s^+$ decays.
We calculate the significances of the combined results from 
$\acpx=0$ by dividing the central values by their 
total uncertainties. These significances are listed in the right-most column 
of table~\ref{tab:FinalAcp} and are mostly less than $1\sigma$. The largest 
significance (for $D^+$, $X=C_{\rm QP}\cos\theta_{\KS}\cos\theta_{\Km}$) 
is $2.5\sigma$, which is plausible as a statistical fluctuation.

%\section{Summary}
%\label{sec:summary}

In summary, we have performed the
first search for $\CP$ violation in four-body $\DDsDecay$ decays
using triple and quadruple products.
We use both Belle and Belle~II datasets corresponding 
to a total integrated luminosity of 1408~$\invfb$. We 
have measured $\CP$ asymmetries 
for six different observables consisting of the triple product $C_{\rm TP}$, the 
quadruple product $C_{\rm QP}$, the product $C_{\rm TP}C_{\rm QP}$, the product of 
the cosines of helicity angles $\theta^{}_{\KS}$ and $\theta^{}_{K^-}$, and the 
products of $C_{\rm TP}$ and $C_{\rm QP}$ with the product  
$\cos\theta^{}_{\KS}\cos\theta^{}_{K^-}$. 
All results are listed in table~\ref{tab:FinalAcp}. 
No evidence for $\CP$ violation is found. These results 
represent the world's most precise measurements of the 
triple-product asymmetry for $D^+_{s}$ decays and for 
singly Cabibbo-suppressed
$D^+$ decays, and the first use of the other $\acpx$ 
asymmetries to search for $\CP$ violation in the charm sector.

\appendix
\section{Simultaneously fitted distributions for $\DpDsp$ data}
\label{app:MDfit}

\begin{figure}[!hbtp]
  \begin{center}% 
  \begin{overpic}[width=1.0\textwidth]{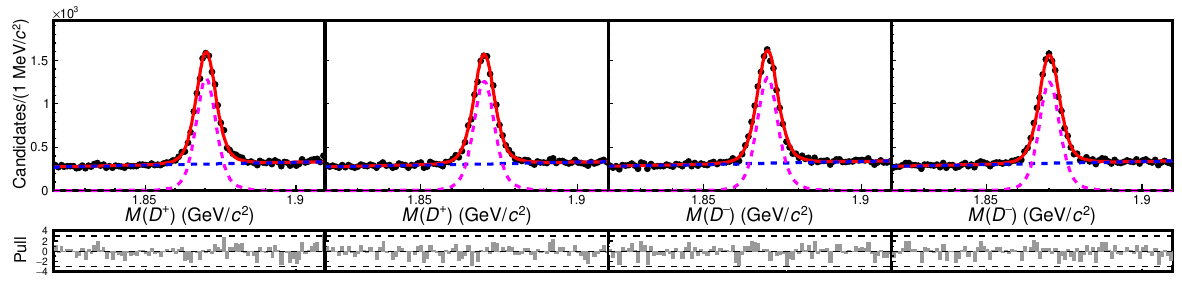}%
  \put(7,16){\footnotesize{$D^+$}}
  \put(7,19){\footnotesize{$C_{\rm TP}>0$}}
  \put(30,16){\footnotesize{$D^+$}}
  \put(30,19){\footnotesize{$C_{\rm TP}<0$}}
  \put(54,16){\footnotesize{$D^-$}}
  \put(54,19){\footnotesize{$-\overline{C}_{\rm TP}>0$}}
  \put(78,16){\footnotesize{$D^-$}}
  \put(78,19){\footnotesize{$-\overline{C}_{\rm TP}<0$}}
  \put(75,23){\footnotesize{{\bf Belle}, $\int\mathcal{L}dt=980~\invfb$}}
  \end{overpic}\\%
  \vskip10pt
  \begin{overpic}[width=1.0\textwidth]{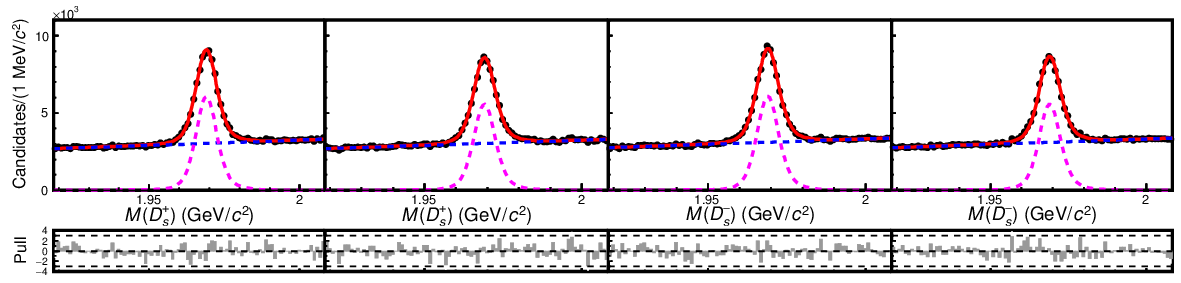}%
  \put(7,16){\footnotesize{$D_s^+$}}
  \put(7,19){\footnotesize{$C_{\rm TP}>0$}}
  \put(30,16){\footnotesize{$D_s^+$}}
  \put(30,19){\footnotesize{$C_{\rm TP}<0$}}
  \put(54,16){\footnotesize{$D_s^-$}}
  \put(54,19){\footnotesize{$-\overline{C}_{\rm TP}>0$}}
  \put(78,16){\footnotesize{$D_s^-$}}
  \put(78,19){\footnotesize{$-\overline{C}_{\rm TP}<0$}}
  \put(75,23){\footnotesize{{\bf Belle}, $\int\mathcal{L}dt=980~\invfb$}}
  \end{overpic}\\%
  \vskip10pt
  \begin{overpic}[width=1.0\textwidth]{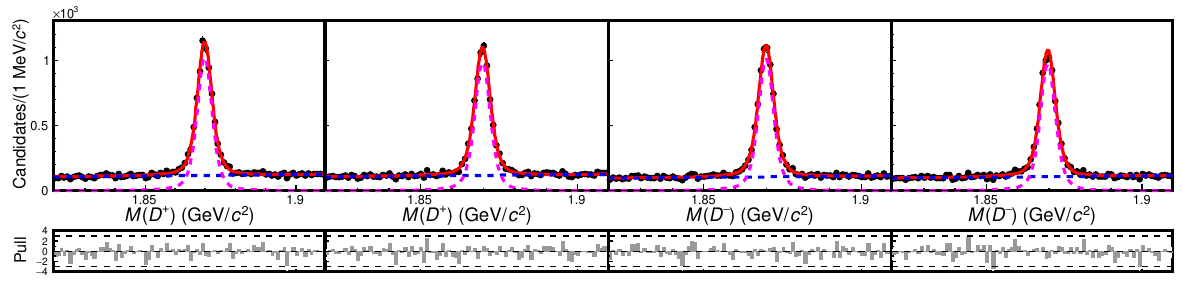}%
  \put(7,16){\footnotesize{$D^+$}}
  \put(7,19){\footnotesize{$C_{\rm TP}>0$}}
  \put(30,16){\footnotesize{$D^+$}}
  \put(30,19){\footnotesize{$C_{\rm TP}<0$}}
  \put(54,16){\footnotesize{$D^-$}}
  \put(54,19){\footnotesize{$-\overline{C}_{\rm TP}>0$}}
  \put(78,16){\footnotesize{$D^-$}}
  \put(78,19){\footnotesize{$-\overline{C}_{\rm TP}<0$}}
  \put(73,23){\footnotesize{{\bf Belle~II}, $\int\mathcal{L}dt=427~\invfb$}}
  \end{overpic}\\%
  \vskip10pt
  \begin{overpic}[width=1.0\textwidth]{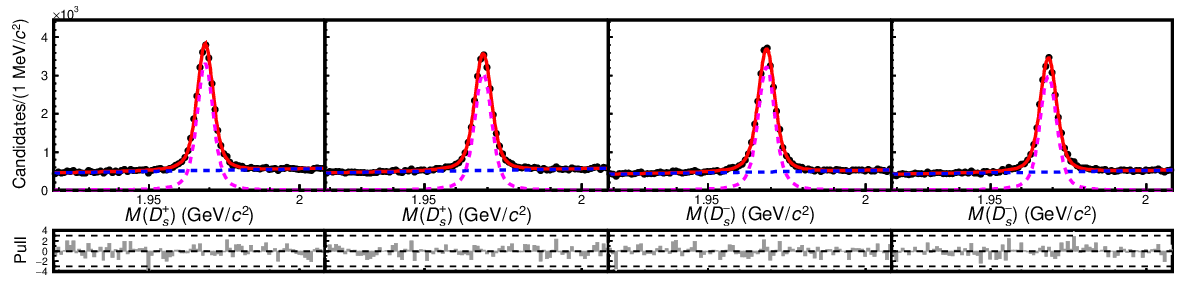}%
  \put(7,16){\footnotesize{$D_s^+$}}
  \put(7,19){\footnotesize{$C_{\rm TP}>0$}}
  \put(30,16){\footnotesize{$D_s^+$}}
  \put(30,19){\footnotesize{$C_{\rm TP}<0$}}
  \put(54,16){\footnotesize{$D_s^-$}}
  \put(54,19){\footnotesize{$-\overline{C}_{\rm TP}>0$}}
  \put(78,16){\footnotesize{$D_s^-$}}
  \put(78,19){\footnotesize{$-\overline{C}_{\rm TP}<0$}}
  \put(73,23){\footnotesize{{\bf Belle~II}, $\int\mathcal{L}dt=427~\invfb$}}
  \end{overpic}
  \vskip-12pt
  \caption{\label{fig:simFit_X1}
    Fitted distributions of $D_{(s)}^\pm\to\KS\Kmp\pi^{\pm}\pi^{\pm}$ 
    data at Belle (top two rows) and Belle~II (bottom two rows), 
    for the triple product $C_{\rm TP}$. 
    The red dashed curves show the fitted signal, and the blue dash-dotted
    curves show the fitted background. The smaller panels
    show the pull distributions.}
  \end{center}
\end{figure}

\begin{figure}[!hbtp]
  \begin{center}% 
  \begin{overpic}[width=1.0\textwidth]{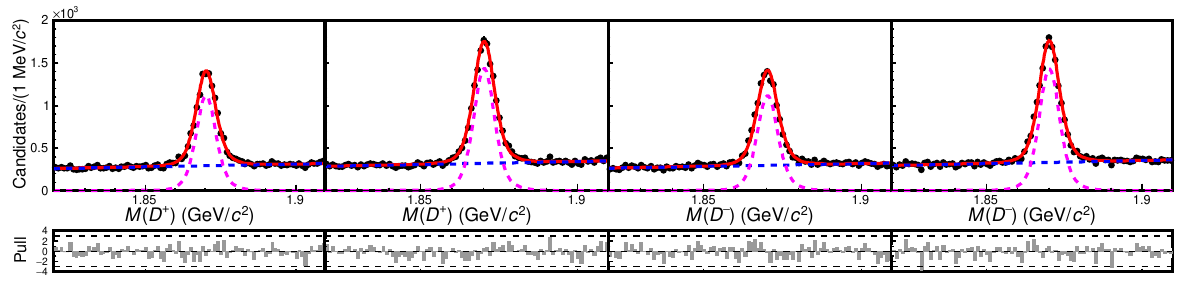}%
  \put(7,16){\footnotesize{$D^+$}}
  \put(7,19){\footnotesize{$C_{\rm QP}>0$}}
  \put(30,16){\footnotesize{$D^+$}}
  \put(30,19){\footnotesize{$C_{\rm QP}<0$}}
  \put(54,16){\footnotesize{$D^-$}}
  \put(54,19){\footnotesize{$\overline{C}_{\rm QP}>0$}}
  \put(78,16){\footnotesize{$D^-$}}
  \put(78,19){\footnotesize{$\overline{C}_{\rm QP}<0$}} 
  \put(75,23){\footnotesize{{\bf Belle}, $\int\mathcal{L}dt=980~\invfb$}}
  \end{overpic}\\%
  \vskip10pt
  \begin{overpic}[width=1.0\textwidth]{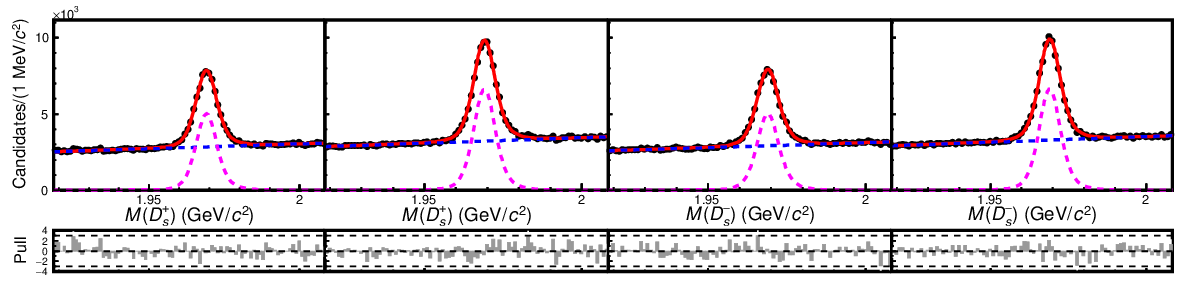}%
  \put(7,16){\footnotesize{$D_s^+$}}
  \put(7,19){\footnotesize{$C_{\rm QP}>0$}}
  \put(30,16){\footnotesize{$D_s^+$}}
  \put(30,19){\footnotesize{$C_{\rm QP}<0$}}
  \put(54,16){\footnotesize{$D_s^-$}}
  \put(54,19){\footnotesize{$\overline{C}_{\rm QP}>0$}}
  \put(78,16){\footnotesize{$D_s^-$}}
  \put(78,19){\footnotesize{$\overline{C}_{\rm QP}<0$}} 
  \put(75,23){\footnotesize{{\bf Belle}, $\int\mathcal{L}dt=980~\invfb$}}
  \end{overpic}\\%
  \vskip10pt
  \begin{overpic}[width=1.0\textwidth]{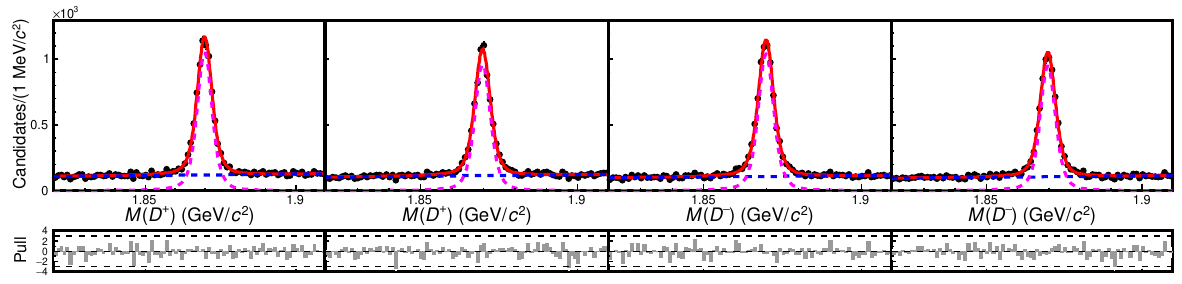}%
  \put(7,16){\footnotesize{$D^+$}}
  \put(7,19){\footnotesize{$C_{\rm QP}>0$}}
  \put(30,16){\footnotesize{$D^+$}}
  \put(30,19){\footnotesize{$C_{\rm QP}<0$}}
  \put(54,16){\footnotesize{$D^-$}}
  \put(54,19){\footnotesize{$\overline{C}_{\rm QP}>0$}}
  \put(78,16){\footnotesize{$D^-$}}
  \put(78,19){\footnotesize{$\overline{C}_{\rm QP}<0$}} 
  \put(73,23){\footnotesize{{\bf Belle~II}, $\int\mathcal{L}dt=427~\invfb$}}
  \end{overpic}\\%  
  \vskip10pt
  \begin{overpic}[width=1.0\textwidth]{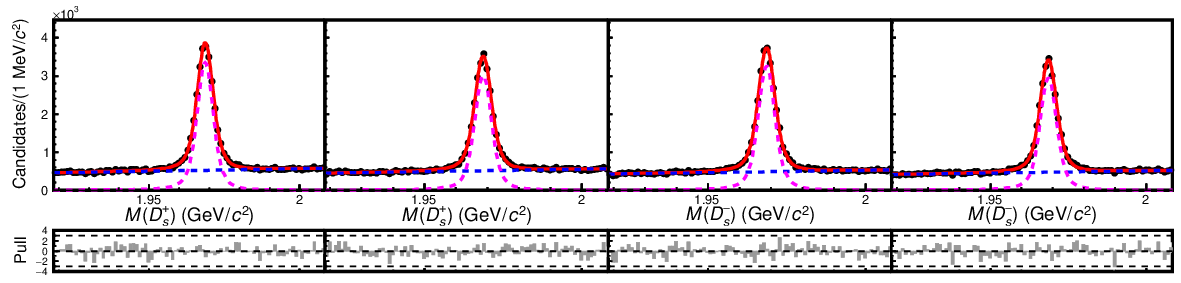}%
  \put(7,16){\footnotesize{$D_s^+$}}
  \put(7,19){\footnotesize{$C_{\rm QP}>0$}}
  \put(30,16){\footnotesize{$D_s^+$}}
  \put(30,19){\footnotesize{$C_{\rm QP}<0$}}
  \put(54,16){\footnotesize{$D_s^-$}}
  \put(54,19){\footnotesize{$\overline{C}_{\rm QP}>0$}}
  \put(78,16){\footnotesize{$D_s^-$}}
  \put(78,19){\footnotesize{$\overline{C}_{\rm QP}<0$}} 
  \put(73,23){\footnotesize{{\bf Belle~II}, $\int\mathcal{L}dt=427~\invfb$}}
  \end{overpic}
  \vskip-12pt
  \caption{\label{fig:simFit_X2}
    Fitted distributions of $D_{(s)}^\pm\to\KS\Kmp\pi^{\pm}\pi^{\pm}$ 
    data at Belle (top two rows) and Belle~II (bottom two rows), 
    for the quadruple product $C_{\rm QP}$. 
    The red dashed curves show the fitted signal, and the blue dash-dotted 
    curves show the fitted background. The smaller panels 
    show the pull distributions.}
  \end{center}
\end{figure}

\begin{figure}[!hbtp]
  \begin{center}% 
  \begin{overpic}[width=1.0\textwidth]{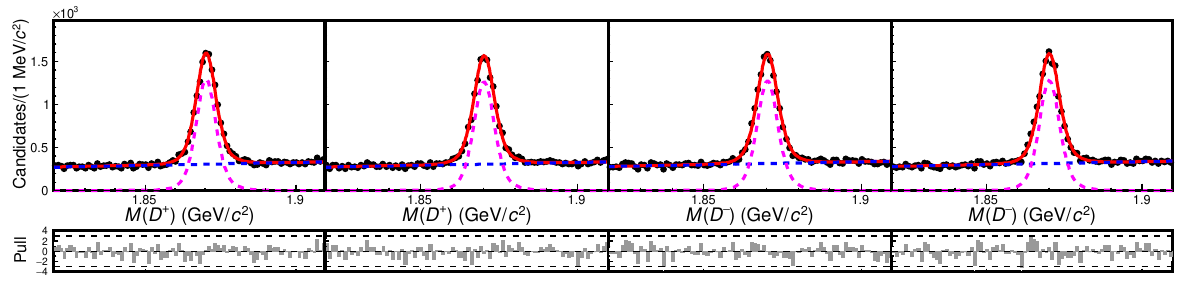}%
  \put(7,16){\footnotesize{$D^+$}}
  \put(7,19){\footnotesize{$X>0$}}
  \put(30,16){\footnotesize{$D^+$}}
  \put(30,19){\footnotesize{$X<0$}}
  \put(54,16){\footnotesize{$D^-$}}
  \put(54,19){\footnotesize{$\overline{X}>0$}}
%  \put(54,19){\footnotesize{$-\overline{C}_{\rm TP}\overline{C}_{\rm QP}>0$}}
  \put(78,16){\footnotesize{$D^-$}}
  \put(78,19){\footnotesize{$\overline{X}<0$}} 
 % \put(78,19){\footnotesize{$-\overline{C}_{\rm TP}\overline{C}_{\rm QP}<0$}} 
  \put(75,23){\footnotesize{{\bf Belle}, $\int\mathcal{L}dt=980~\invfb$}}
  \end{overpic}\\%
  \vskip10pt
  \begin{overpic}[width=1.0\textwidth]{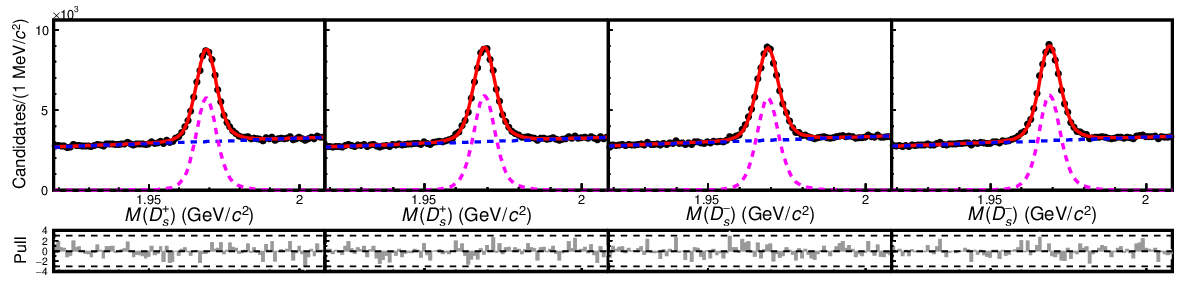}%
  \put(7,16){\footnotesize{$D_s^+$}}
  \put(7,19){\footnotesize{$X>0$}}
  \put(30,16){\footnotesize{$D_s^+$}}
  \put(30,19){\footnotesize{$X<0$}}
  \put(54,16){\footnotesize{$D_s^-$}}
  \put(54,19){\footnotesize{$\overline{X}>0$}}
  \put(78,16){\footnotesize{$D_s^-$}}
  \put(78,19){\footnotesize{$\overline{X}<0$}} 
  \put(75,23){\footnotesize{{\bf Belle}, $\int\mathcal{L}dt=980~\invfb$}}
  \end{overpic}\\%
  \vskip10pt
  \begin{overpic}[width=1.0\textwidth]{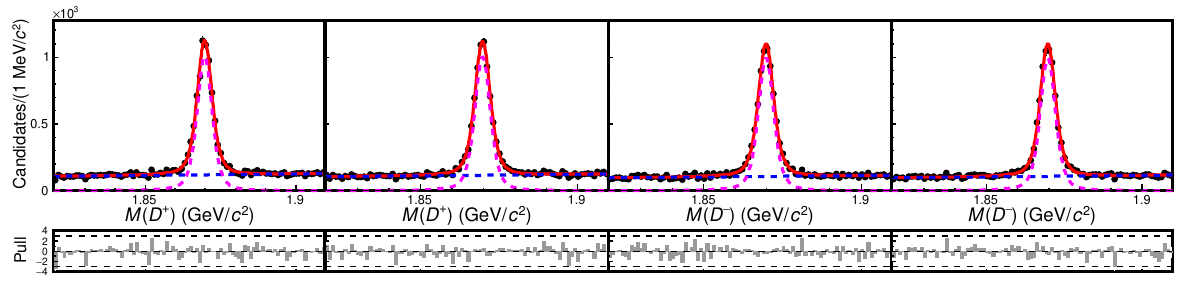}%
  \put(7,16){\footnotesize{$D^+$}}
  \put(7,19){\footnotesize{$X>0$}}
  \put(30,16){\footnotesize{$D^+$}}
  \put(30,19){\footnotesize{$X<0$}}
  \put(54,16){\footnotesize{$D^-$}}
  \put(54,19){\footnotesize{$\overline{X}>0$}}
  \put(78,16){\footnotesize{$D^-$}}
  \put(78,19){\footnotesize{$\overline{X}<0$}} 
  \put(73,23){\footnotesize{{\bf Belle~II}, $\int\mathcal{L}dt=427~\invfb$}}
  \end{overpic}\\%  
  \vskip10pt
  \begin{overpic}[width=1.0\textwidth]{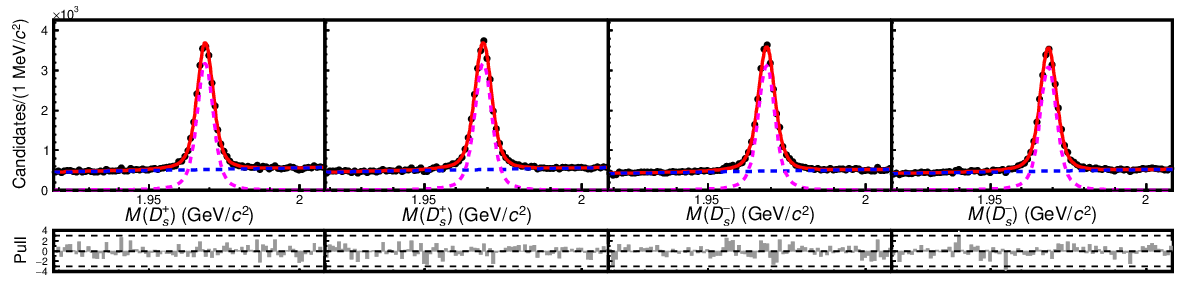}%
  \put(7,16){\footnotesize{$D_s^+$}}
  \put(7,19){\footnotesize{$X>0$}}
  \put(30,16){\footnotesize{$D_s^+$}}
  \put(30,19){\footnotesize{$X<0$}}
  \put(54,16){\footnotesize{$D_s^-$}}
  \put(54,19){\footnotesize{$\overline{X}>0$}}
  \put(78,16){\footnotesize{$D_s^-$}}
  \put(78,19){\footnotesize{$\overline{X}<0$}} 
  \put(73,23){\footnotesize{{\bf Belle~II}, $\int\mathcal{L}dt=427~\invfb$}}
  \end{overpic}\\ 
  \vskip-12pt
  \caption{\label{fig:simFit_X3}
    Fitted distributions of $D_{(s)}^\pm\to\KS\Kmp\pi^{\pm}\pi^{\pm}$ 
    data at Belle (top two rows) and Belle~II (bottom two rows) for
    $X=C_{\rm TP}\,C_{\rm QP}$.
    The red dashed curves show the fitted signal, and the blue 
    dash-dotted curves show the fitted background. The smaller panels
    show the pull distributions.}
  \end{center}
\end{figure}

\begin{figure}[!hbtp]
  \begin{center}% 
  \begin{overpic}[width=1.0\textwidth]{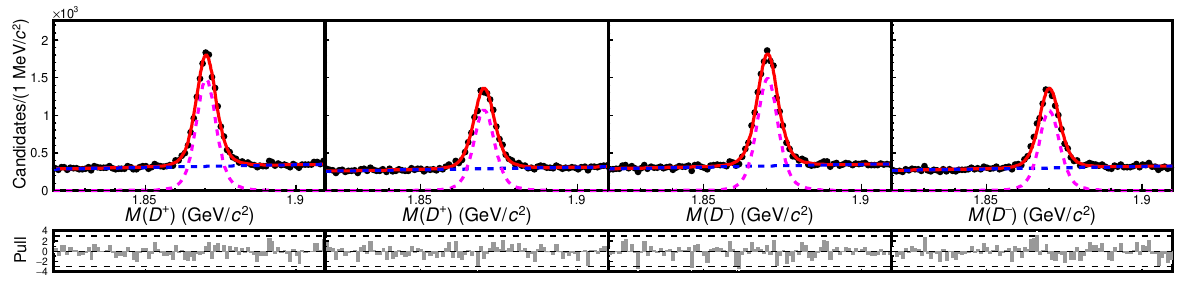}%
  \put(7,16){\footnotesize{$D^+$}}
  \put(7,19){\footnotesize{$X>0$}}
  \put(30,16){\footnotesize{$D^+$}}
  \put(30,19){\footnotesize{$X<0$}}
  \put(54,16){\footnotesize{$D^-$}}
  \put(54,19){\footnotesize{$\overline{X}>0$}}
  \put(78,16){\footnotesize{$D^-$}}
  \put(78,19){\footnotesize{$\overline{X}<0$}} 
  \put(75,23){\footnotesize{{\bf Belle}, $\int\mathcal{L}dt=980~\invfb$}}
  \end{overpic}\\%
  \vskip10pt
  \begin{overpic}[width=1.0\textwidth]{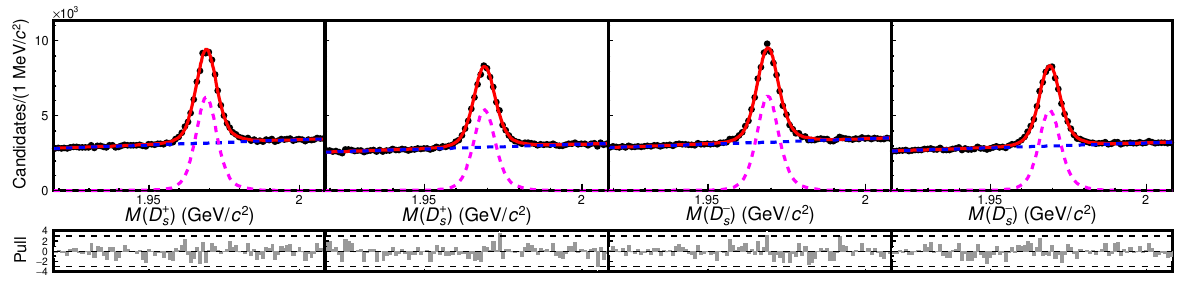}%
  \put(7,16){\footnotesize{$D_s^+$}}
  \put(7,19){\footnotesize{$X>0$}}
  \put(30,16){\footnotesize{$D_s^+$}}
  \put(30,19){\footnotesize{$X<0$}}
  \put(54,16){\footnotesize{$D_s^-$}}
  \put(54,19){\footnotesize{$\overline{X}>0$}}
  \put(78,16){\footnotesize{$D_s^-$}}
  \put(78,19){\footnotesize{$\overline{X}<0$}} 
  \put(75,23){\footnotesize{{\bf Belle}, $\int\mathcal{L}dt=980~\invfb$}}
  \end{overpic}\\%
  \vskip10pt
  \begin{overpic}[width=1.0\textwidth]{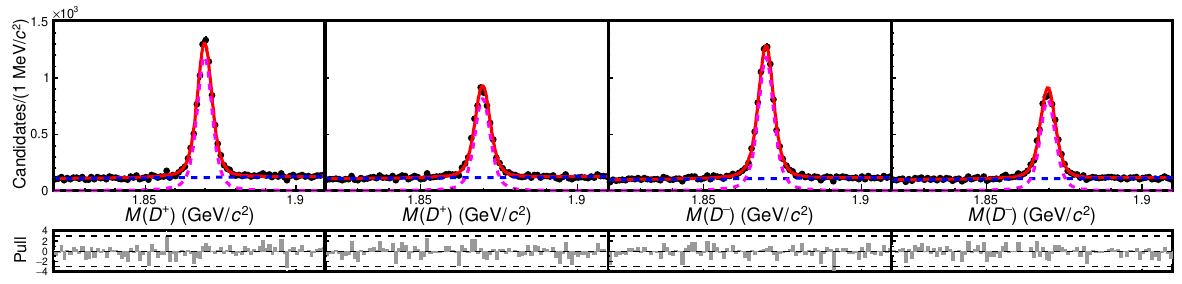}%
  \put(7,16){\footnotesize{$D^+$}}
  \put(7,19){\footnotesize{$X>0$}}
  \put(30,16){\footnotesize{$D^+$}}
  \put(30,19){\footnotesize{$X<0$}}
  \put(54,16){\footnotesize{$D^-$}}
  \put(54,19){\footnotesize{$\overline{X}>0$}}
  \put(78,16){\footnotesize{$D^-$}}
  \put(78,19){\footnotesize{$\overline{X}<0$}} 
  \put(73,23){\footnotesize{{\bf Belle~II}, $\int\mathcal{L}dt=427~\invfb$}}
  \end{overpic}\\%
  \vskip10pt  
  \begin{overpic}[width=1.0\textwidth]{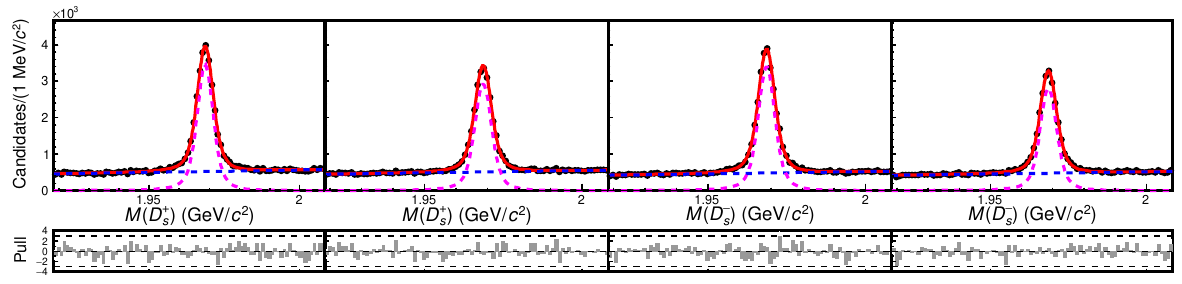}%
  \put(7,16){\footnotesize{$D_s^+$}}
  \put(7,19){\footnotesize{$X>0$}}
  \put(30,16){\footnotesize{$D_s^+$}}
  \put(30,19){\footnotesize{$X<0$}}
  \put(54,16){\footnotesize{$D_s^-$}}
  \put(54,19){\footnotesize{$\overline{X}>0$}}
  \put(78,16){\footnotesize{$D_s^-$}}
  \put(78,19){\footnotesize{$\overline{X}<0$}} 
  \put(73,23){\footnotesize{{\bf Belle~II}, $\int\mathcal{L}dt=427~\invfb$}}
  \end{overpic}  
  \vskip-12pt
  \caption{\label{fig:simFit_X4}
    Fitted distributions of $D_{(s)}^\pm\to\KS\Kmp\pi^{\pm}\pi^{\pm}$ data at Belle (top two rows) and Belle~II (bottom two rows) for $X=\cos\theta_{\KS}\cos\theta_{\Km}$.
    The red dashed curves show the fitted signal, and the blue 
    dash-dotted curves show the fitted background. The smaller panels
    show the pull distributions.}
  \end{center}
\end{figure}

\begin{figure}[!hbtp]
  \begin{center}% 
  \begin{overpic}[width=1.0\textwidth]{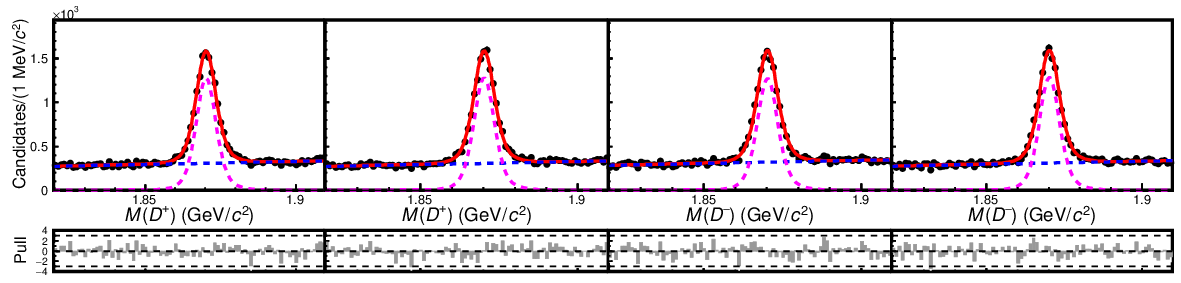}%
  \put(7,16){\footnotesize{$D^+$}}
  \put(7,19){\footnotesize{$X>0$}}
  \put(30,16){\footnotesize{$D^+$}}
  \put(30,19){\footnotesize{$X<0$}}
  \put(54,16){\footnotesize{$D^-$}}
  \put(54,19){\footnotesize{$\overline{X}>0$}}
  \put(78,16){\footnotesize{$D^-$}}
  \put(78,19){\footnotesize{$\overline{X}<0$}}  
  \put(75,23){\footnotesize{{\bf Belle}, $\int\mathcal{L}dt=980~\invfb$}}
  \end{overpic}\\%
  \vskip10pt
  \begin{overpic}[width=1.0\textwidth]{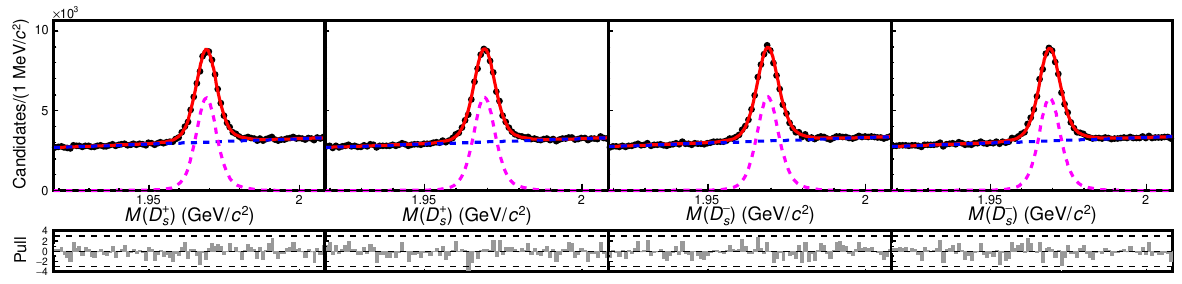}%
  \put(7,16){\footnotesize{$D_s^+$}}
  \put(7,19){\footnotesize{$X>0$}}
  \put(30,16){\footnotesize{$D_s^+$}}
  \put(30,19){\footnotesize{$X<0$}}
  \put(54,16){\footnotesize{$D_s^-$}}
  \put(54,19){\footnotesize{$\overline{X}>0$}}
  \put(78,16){\footnotesize{$D_s^-$}}
  \put(78,19){\footnotesize{$\overline{X}<0$}}  
  \put(75,23){\footnotesize{{\bf Belle}, $\int\mathcal{L}dt=980~\invfb$}}
  \end{overpic}\\%  
  \vskip10pt
  \begin{overpic}[width=1.0\textwidth]{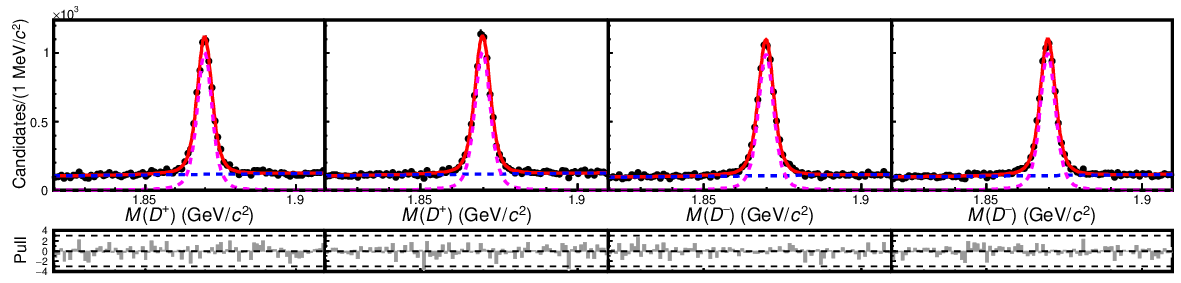}%
  \put(7,16){\footnotesize{$D^+$}}
  \put(7,19){\footnotesize{$X>0$}}
  \put(30,16){\footnotesize{$D^+$}}
  \put(30,19){\footnotesize{$X<0$}}
  \put(54,16){\footnotesize{$D^-$}}
  \put(54,19){\footnotesize{$\overline{X}>0$}}
  \put(78,16){\footnotesize{$D^-$}}
  \put(78,19){\footnotesize{$\overline{X}<0$}}  
  \put(73,23){\footnotesize{{\bf Belle~II}, $\int\mathcal{L}dt=427~\invfb$}}
  \end{overpic}\\%
  \vskip10pt
  \begin{overpic}[width=1.0\textwidth]{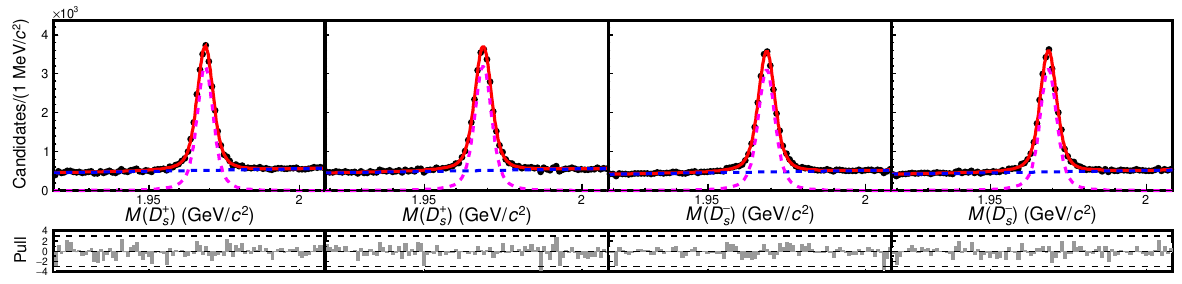}%
  \put(7,16){\footnotesize{$D_s^+$}}
  \put(7,19){\footnotesize{$X>0$}}
  \put(30,16){\footnotesize{$D_s^+$}}
  \put(30,19){\footnotesize{$X<0$}}
  \put(54,16){\footnotesize{$D_s^-$}}
  \put(54,19){\footnotesize{$\overline{X}>0$}}
  \put(78,16){\footnotesize{$D_s^-$}}
  \put(78,19){\footnotesize{$\overline{X}<0$}}  
  \put(73,23){\footnotesize{{\bf Belle~II}, $\int\mathcal{L}dt=427~\invfb$}}
  \end{overpic} 
  \vskip-12pt
  \caption{\label{fig:simFit_X5}
    Fitted distributions of $D^\pm\to\KS\Kmp\pi^{\pm}\pi^{\pm}$ data at Belle (top two rows) and Belle~II (bottom two rows) for 
    $X=C_{\rm TP}\cos\theta_{\KS}\cos\theta_{\Km}$.
    The red dashed curves show the fitted signal, and the blue 
    dash-dotted curves show the fitted background. The smaller panels
    show the pull distributions.}
  \end{center}
\end{figure}

\begin{figure}[!hbtp]
  \begin{center}% 
  \begin{overpic}[width=1.0\textwidth]{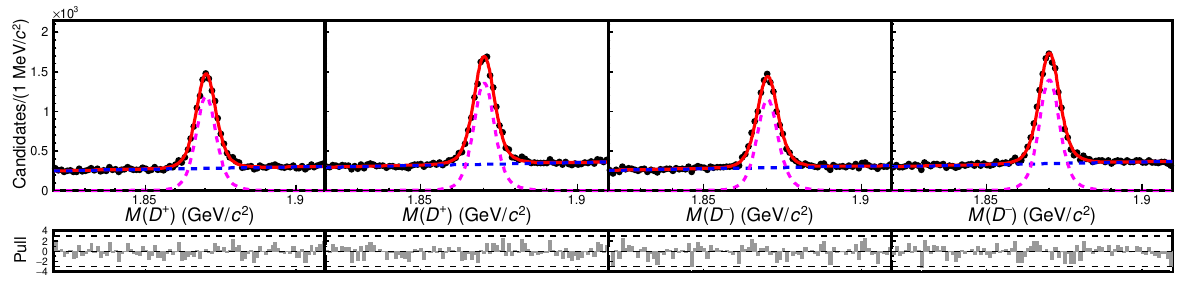}%
  \put(7,16){\footnotesize{$D^+$}}
  \put(7,19){\footnotesize{$X>0$}}
  \put(30,16){\footnotesize{$D^+$}}
  \put(30,19){\footnotesize{$X<0$}}
  \put(54,16){\footnotesize{$D^-$}}
  \put(54,19){\footnotesize{$\overline{X}>0$}}
  \put(78,16){\footnotesize{$D^-$}}
  \put(78,19){\footnotesize{$\overline{X}<0$}} 
  \put(75,23){\footnotesize{{\bf Belle}, $\int\mathcal{L}dt=980~\invfb$}}
  \end{overpic}\\%
  \vskip10pt
  \begin{overpic}[width=1.0\textwidth]{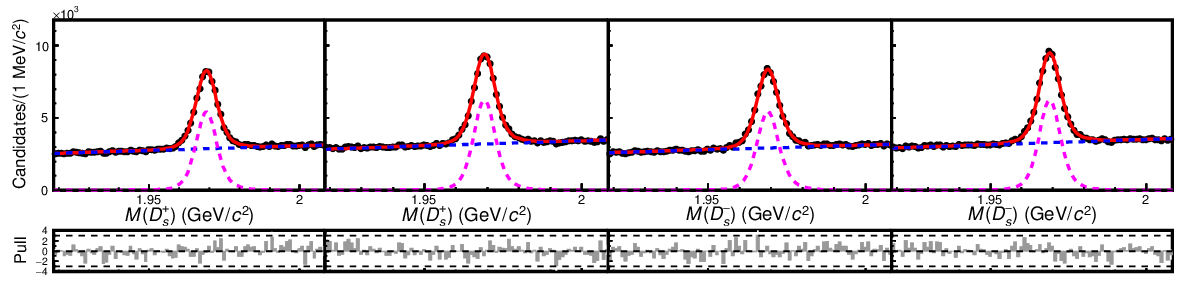}%
  \put(7,16){\footnotesize{$D_s^+$}}
  \put(7,19){\footnotesize{$X>0$}}
  \put(30,16){\footnotesize{$D_s^+$}}
  \put(30,19){\footnotesize{$X<0$}}
  \put(54,16){\footnotesize{$D_s^-$}}
  \put(54,19){\footnotesize{$\overline{X}>0$}}
  \put(78,16){\footnotesize{$D_s^-$}}
  \put(78,19){\footnotesize{$\overline{X}<0$}} 
  \put(75,23){\footnotesize{{\bf Belle}, $\int\mathcal{L}dt=980~\invfb$}}
  \end{overpic}\\%
  \vskip10pt
  \begin{overpic}[width=1.0\textwidth]{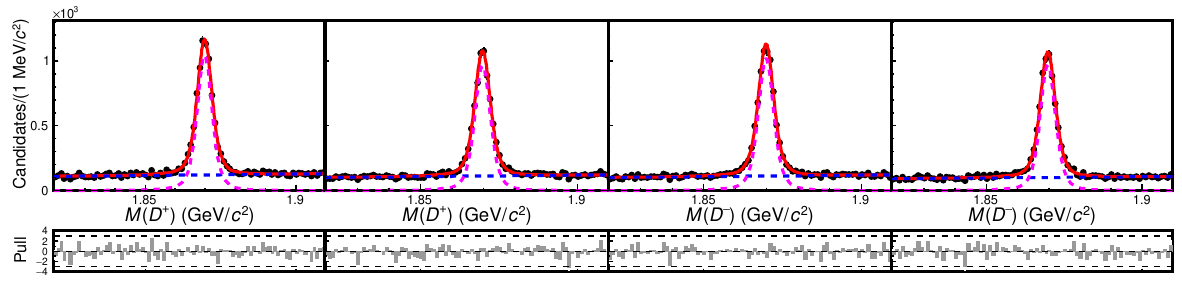}%
  \put(7,16){\footnotesize{$D^+$}}
  \put(7,19){\footnotesize{$X>0$}}
  \put(30,16){\footnotesize{$D^+$}}
  \put(30,19){\footnotesize{$X<0$}}
  \put(54,16){\footnotesize{$D^-$}}
  \put(54,19){\footnotesize{$\overline{X}>0$}}
  \put(78,16){\footnotesize{$D^-$}}
  \put(78,19){\footnotesize{$\overline{X}<0$}} 
  \put(73,23){\footnotesize{{\bf Belle~II}, $\int\mathcal{L}dt=427~\invfb$}}
  \end{overpic}\\%  
  \vskip10pt
  \begin{overpic}[width=1.0\textwidth]{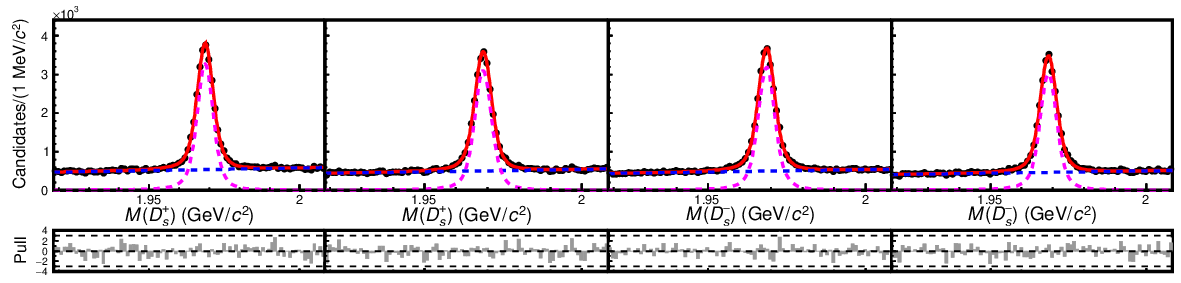}%
  \put(7,16){\footnotesize{$D_s^+$}}
  \put(7,19){\footnotesize{$X>0$}}
  \put(30,16){\footnotesize{$D_s^+$}}
  \put(30,19){\footnotesize{$X<0$}}
  \put(54,16){\footnotesize{$D_s^-$}}
  \put(54,19){\footnotesize{$\overline{X}>0$}}
  \put(78,16){\footnotesize{$D_s^-$}}
  \put(78,19){\footnotesize{$\overline{X}<0$}} 
  \put(73,23){\footnotesize{{\bf Belle~II}, $\int\mathcal{L}dt=427~\invfb$}}
  \end{overpic}  
  \vskip-12pt
  \caption{\label{fig:simFit_X6}
    Fitted distributions of $D^\pm\to\KS\Kmp\pi^{\pm}\pi^{\pm}$ data at Belle (top two rows) and Belle~II (bottom two rows) for 
    $X=C_{\rm QP}\cos\theta_{\KS}\cos\theta_{\Km}$.
    The red dashed curves show the fitted signal, and the blue 
    dash-dotted curves show the fitted background. The smaller panels
    show the pull distributions.}
  \end{center}
\end{figure}

The $M(D)$ distributions for $D^+_{(s)}\to\KS\Kmp\pi^{\pm}\pi^{\pm}$ 
decays used to fit for the asymmetries $\acpx$, %along with projections of the fit results
with fit results overlaid, are shown in 
figs.~\ref{fig:simFit_X1}-\ref{fig:simFit_X6}. Specifically, 
fig.~\ref{fig:simFit_X1} corresponds to $X=C_{\rm TP}$, 
fig.~\ref{fig:simFit_X2} to $X=C_{\rm QP}$, 
fig.~\ref{fig:simFit_X3} to $X=C_{\rm TP}\,C_{\rm QP}$, 
fig.~\ref{fig:simFit_X4} to $X=\cos\theta_{\KS}\cos\theta_{\Km}$, 
fig.~\ref{fig:simFit_X5} to $X=C_{\rm TP}\cos\theta_{\KS}\cos\theta_{\Km}$, and fig.~\ref{fig:simFit_X6} to 
$X=C_{\rm QP}\cos\theta_{\KS}\cos\theta_{\Km}$. 

%\cleardoublepage

\section{Triple- and quadruple-product distributions}
\label{app:Xplots}

\begin{figure}[!hbtp]
  \begin{center}% 
%  \begin{overpic}[width=0.5\textwidth]{CTP_DpSCS_data}%  
  \begin{overpic}[width=0.5\textwidth]{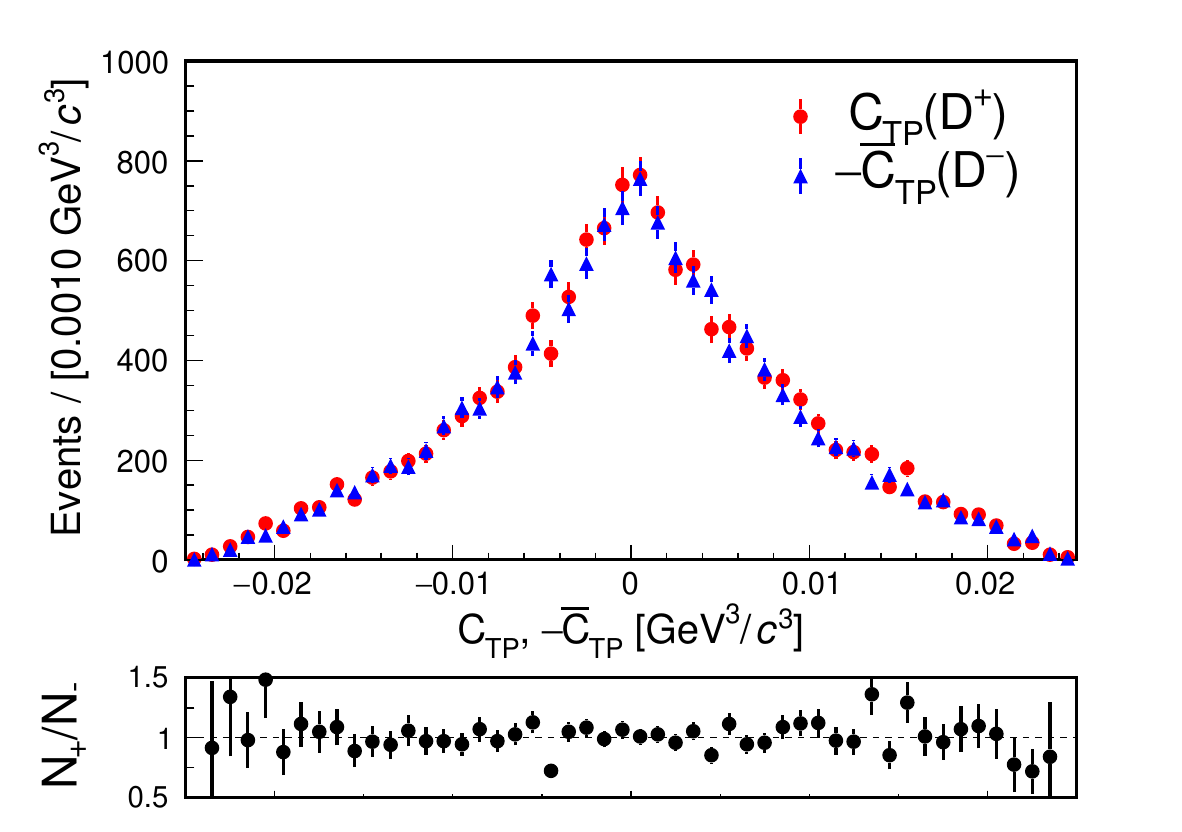}%  
  \put(18,58){\footnotesize{$D^{\pm}\to\KS\Kmp\pi^{\pm}\pi^{\pm}$}}
  \end{overpic}%
%  \begin{overpic}[width=0.5\textwidth]{CQP_DpSCS_data}%  
  \begin{overpic}[width=0.5\textwidth]{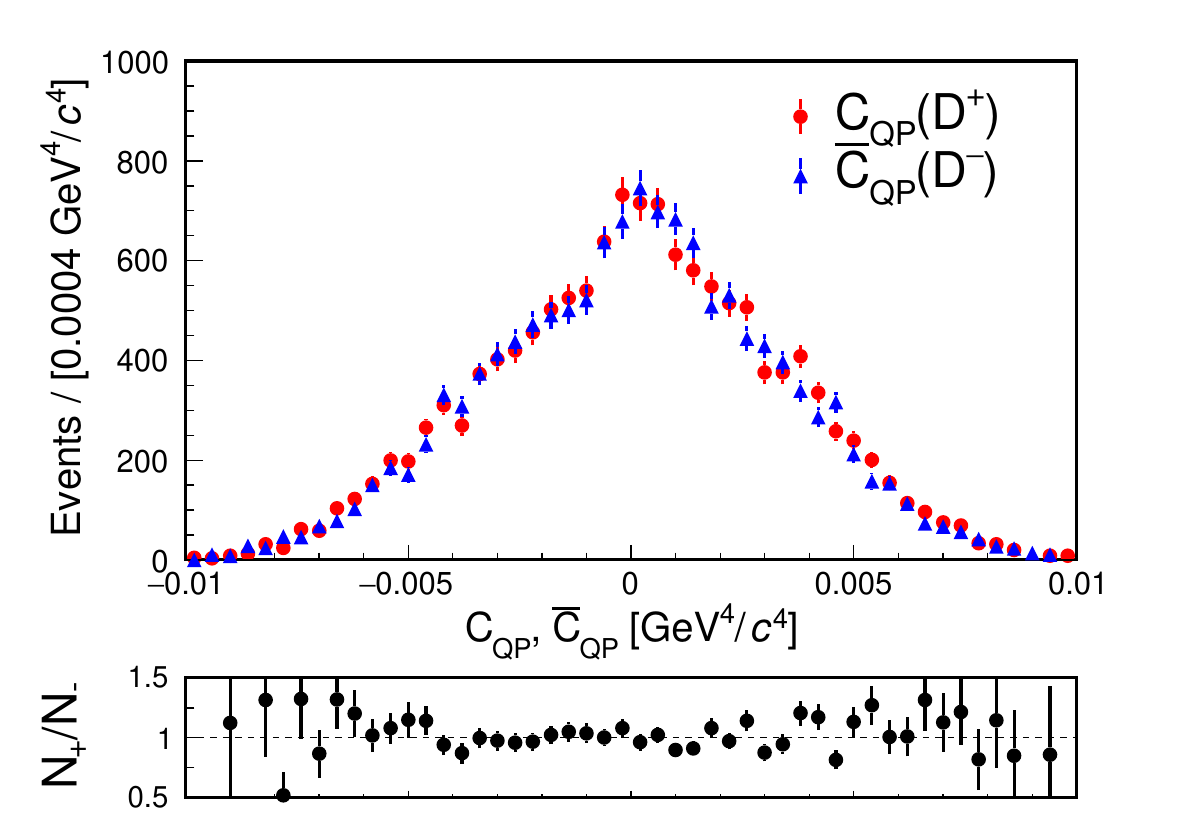}%  
  \put(18,58){\footnotesize{$D^{\pm}\to\KS\Kmp\pi^{\pm}\pi^{\pm}$}}
  \put(26,67){\footnotesize{{\bf Belle~II} data, $\int\mathcal{L}dt=427~\invfb$}}
  \end{overpic}\\
%  \vskip5pt
%  \begin{overpic}[width=0.5\textwidth]{CTP_DsCF_data}%  
  \begin{overpic}[width=0.5\textwidth]{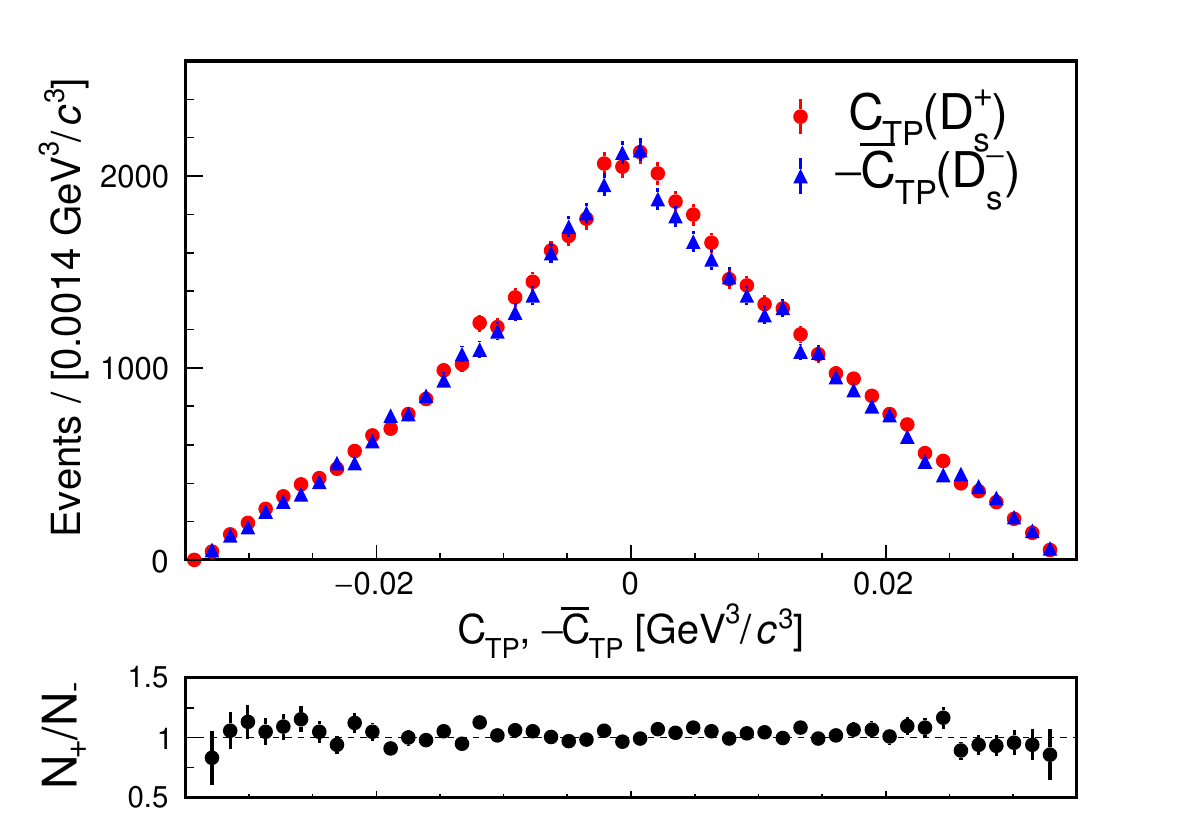}%  
  \put(18,58){\footnotesize{$D_{s}^{\pm}\to\KS\Kmp\pi^{\pm}\pi^{\pm}$}}
  \end{overpic}%
%  \begin{overpic}[width=0.5\textwidth]{CQP_DsCF_data}%  
  \begin{overpic}[width=0.5\textwidth]{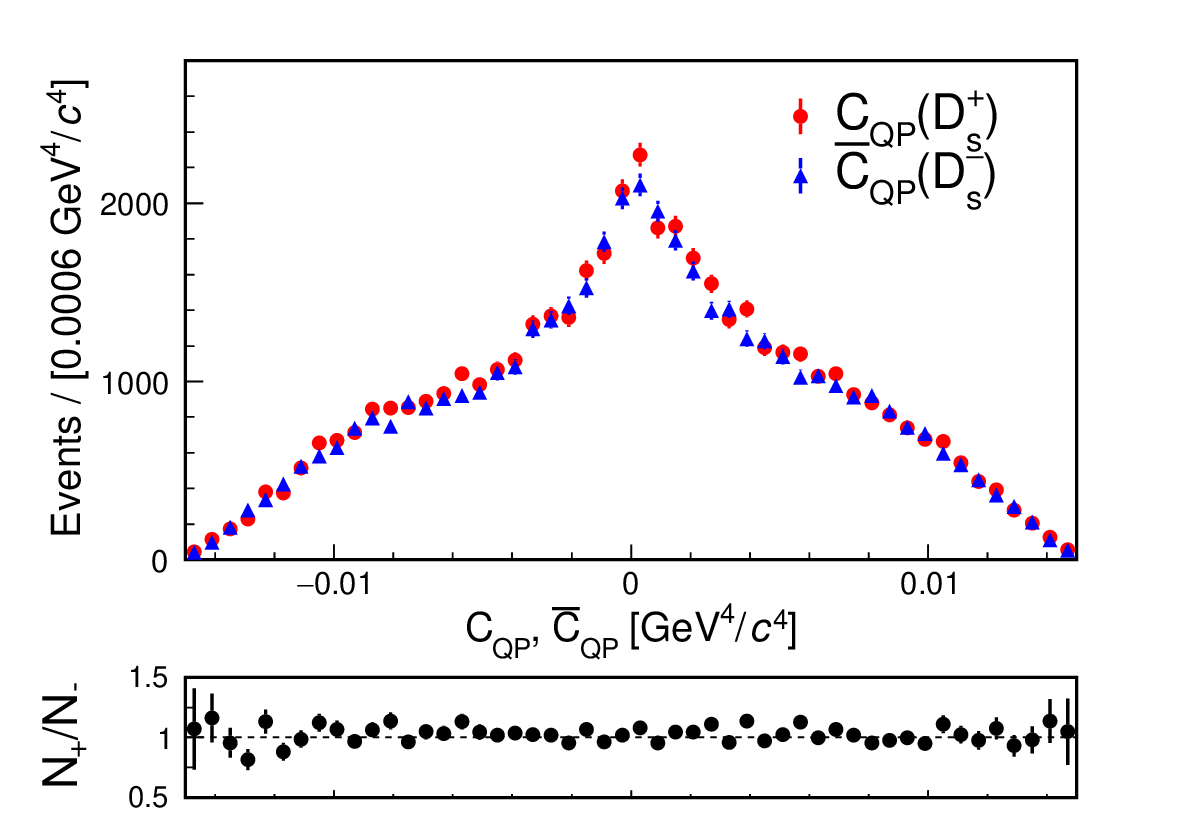}%  
  \put(18,58){\footnotesize{$D_{s}^{\pm}\to\KS\Kmp\pi^{\pm}\pi^{\pm}$}}
%  \put(26,67){\footnotesize{{\bf Belle~II} data, $\int\mathcal{L}dt=427~\invfb$}}
  \end{overpic}    
%  \vskip-10pt
  \caption{\label{fig:Xplots}
Belle~II distributions for $C_{\rm TP}$ (left) and $C_{\rm QP}$ (right) for
$D_{(s)}^{+}\to\KS\Km\pi^{+}\pi^{+}$ candidates, and 
$-\overline{C}_{\rm TP}$ (left) and $\overline{C}_{\rm QP}$ (right) 
for $D_{(s)}^{-}\to\KS\Kp\pi^{-}\pi^{-}$ candidates. 
Points with error bars denote candidates in the signal 
region $|M(D)-m_D|<10~{\rm MeV}/c^2$ after background 
(scaled from the sideband 
$20~{\rm MeV}/c^2<|M(D)-m_D|<40~{\rm MeV}/c^2$)
has been subtracted. The lower panels show the ratios
$C^{}_{\rm TP}/(-\overline{C}^{}_{\rm TP})$
and $C^{}_{\rm QP}/\overline{C}^{}_{\rm QP}$.}
%  Distributions of $C_{\rm TP}$ ($D_{(s)}^+$) and $-\overline{C}_{\rm TP}$ ($D_{(s)}^-$), and $C_{\rm QP}$ ($D_{(s)}^+$) and $\overline{C}_{\rm QP}$ ($D_{(s)}^-$) for $D_{(s)}^{\pm}\to\KS\Kmp\pi^{\pm}\pi^{\pm}$ candidates from Belle~II data. 
%  The points with error bars denote candidates in $M(D)$ signal region~(S.R.): $|M(D)-m_D|<10~{\rm MeV}/{c}^2$; and histograms denote the backgrounds estimated by candidates in $M(D)$ sideband~(S.B.): $20~{\rm MeV}/{c}^2<|M(D)-m_D|<40~{\rm MeV}/{c}^2$ (a scaling factor of 0.5 is applied for histograms).}}
  \end{center}
\end{figure}

The Belle~II distributions of the triple product $C_{\rm TP}$
and quadruple product $C_{\rm QP}$ for 
$D_{(s)}^{+}\to\KS\Km\pi^{+}\pi^{+}$ candidates,
and $-\overline{C}_{\rm TP}$ and $\overline{C}_{\rm QP}$ for 
$D_{(s)}^{-}\to\KS\Kp\pi^{-}\pi^{-}$ candidates,
are shown in fig.~\ref{fig:Xplots}.

\acknowledgments
% Policy from October 20, 2022
This work, based on data collected using the Belle II detector, which was built and commissioned prior to March 2019,
and data collected using the Belle detector, which was operated until June 2010,
was supported by
%Armenia
Higher Education and Science Committee of the Republic of Armenia Grant No.~23LCG-1C011;
%Australia
Australian Research Council and Research Grants
No.~DP200101792, % Jackson
No.~DP210101900, % Urquijo
No.~DP210102831, % Sevior
No.~DE220100462, % Hsu
No.~LE210100098, % Infrastructure
and
No.~LE230100085; % Infrastructure
%Austria
Austrian Federal Ministry of Education, Science and Research,
Austrian Science Fund
No.~P~34529,
No.~J~4731,
No.~J~4625,
and
No.~M~3153,
and
Horizon 2020 ERC Starting Grant No.~947006 ``InterLeptons'';
%Canada
Natural Sciences and Engineering Research Council of Canada, Compute Canada and CANARIE;
%China
National Key R\&D Program of China under Contract No.~2022YFA1601903,
National Natural Science Foundation of China and Research Grants
No.~11575017,
No.~11761141009,
No.~11705209,
No.~11975076,
No.~12135005,
No.~12150004,
No.~12161141008,
and
No.~12175041,
and Shandong Provincial Natural Science Foundation Project~ZR2022JQ02;
%Czech Republic
the Czech Science Foundation Grant No.~22-18469S 
and
Charles University Grant Agency project No.~246122;
%EU
European Research Council, Seventh Framework PIEF-GA-2013-622527,
Horizon 2020 ERC-Advanced Grants No.~267104 and No.~884719,
Horizon 2020 ERC-Consolidator Grant No.~819127,
Horizon 2020 Marie Sklodowska-Curie Grant Agreement No.~700525 ``NIOBE''
and
No.~101026516,
and
Horizon 2020 Marie Sklodowska-Curie RISE project JENNIFER2 Grant Agreement No.~822070 (European grants);
%France
L'Institut National de Physique Nucl\'{e}aire et de Physique des Particules (IN2P3) du CNRS
and
L'Agence Nationale de la Recherche (ANR) under grant ANR-21-CE31-0009 (France);
%Germany
BMBF, DFG, HGF, MPG, and AvH Foundation (Germany);
%India
Department of Atomic Energy under Project Identification No.~RTI 4002,
Department of Science and Technology,
and
UPES SEED funding programs
No.~UPES/R\&D-SEED-INFRA/17052023/01 and
No.~UPES/R\&D-SOE/20062022/06 (India);
%Israel
Israel Science Foundation Grant No.~2476/17,
U.S.-Israel Binational Science Foundation Grant No.~2016113, and
Israel Ministry of Science Grant No.~3-16543;
%Italy
Istituto Nazionale di Fisica Nucleare and the Research Grants BELLE2;
%Japan
Japan Society for the Promotion of Science, Grant-in-Aid for Scientific Research Grants
No.~16H03968,
No.~16H03993,
No.~16H06492,
No.~16K05323,
No.~17H01133,
No.~17H05405,
No.~18K03621,
No.~18H03710,
No.~18H05226,
No.~19H00682, % Niigata
No.~20H05850,
No.~20H05858,
No.~22H00144,
No.~22K14056,
No.~22K21347,
No.~23H05433,
No.~26220706,
and
No.~26400255,
%the National Institute of Informatics, and Science Information NETwork 5 (SINET5), 
and
the Ministry of Education, Culture, Sports, Science, and Technology (MEXT) of Japan;  
%Korea
National Research Foundation (NRF) of Korea Grants
No.~2016R1-D1A1B-02012900,
No.~2018R1-A6A1A-06024970,
No.~2021R1-A6A1A-03043957,
No.~2021R1-F1A-1060423,
No.~2021R1-F1A-1064008,
No.~2022R1-A2C-1003993,
No.~2022R1-A2C-1092335,
No.~RS-2023-00208693,
No.~RS-2024-00354342
and
No.~RS-2022-00197659,
Radiation Science Research Institute,
Foreign Large-Size Research Facility Application Supporting project,
the Global Science Experimental Data Hub Center, the Korea Institute of
Science and Technology Information (K24L2M1C4)
and
KREONET/GLORIAD;
%Malaysia
Universiti Malaya RU grant, Akademi Sains Malaysia, and Ministry of Education Malaysia;
%Mexico
% CINVESTAV-IPN, UNAM, UAS, BUAP and CONACYT are funded under
Frontiers of Science Program Contracts
No.~FOINS-296,
No.~CB-221329,
No.~CB-236394,
No.~CB-254409,
and
No.~CB-180023, and SEP-CINVESTAV Research Grant No.~237 (Mexico);
%Poland
the Polish Ministry of Science and Higher Education and the National Science Center;
%Russia
the Ministry of Science and Higher Education of the Russian Federation
and
the HSE University Basic Research Program, Moscow;
%Saudi Arabia
University of Tabuk Research Grants
No.~S-0256-1438 and No.~S-0280-1439 (Saudi Arabia);
%Slovenia
Slovenian Research Agency and Research Grants
No.~J1-9124
and
No.~P1-0135;
%Spain
Ikerbasque, Basque Foundation for Science,
the State Agency for Research of the Spanish Ministry of Science and Innovation through Grant No. PID2022-136510NB-C33,
Agencia Estatal de Investigacion, Spain
Grant No.~RYC2020-029875-I
and
Generalitat Valenciana, Spain
Grant No.~CIDEGENT/2018/020;
%Swiss (Belle 1)
the Swiss National Science Foundation;
%Sweden
The Knut and Alice Wallenberg Foundation (Sweden), Contracts No.~2021.0174 and No.~2021.0299;
%Taiwan
National Science and Technology Council,
and
Ministry of Education (Taiwan);
%Thailand
Thailand Center of Excellence in Physics;
%Turkey
TUBITAK ULAKBIM (Turkey);
%Ukraine
National Research Foundation of Ukraine, Project No.~2020.02/0257,
and
Ministry of Education and Science of Ukraine;
%USA
the U.S. National Science Foundation and Research Grants
No.~PHY-1913789 % Indiana CEEM
and
No.~PHY-2111604, % Luther
and the U.S. Department of Energy and Research Awards
No.~DE-AC06-76RLO1830, % PNNL
No.~DE-SC0007983, % Wayne State
No.~DE-SC0009824, % Florida
No.~DE-SC0009973, % VPI
No.~DE-SC0010007, % Duke
No.~DE-SC0010073, % South Carolina
No.~DE-SC0010118, % Carnegie Mellon
No.~DE-SC0010504, % Hawaii
No.~DE-SC0011784, % Cincinnati
No.~DE-SC0012704, % BNL
No.~DE-SC0019230, % Duke
No.~DE-SC0021274, % Mississippi
No.~DE-SC0021616, % Mississippi
No.~DE-SC0022350, % Louisville
No.~DE-SC0023470; % South Alabama
%last group
and
%Vietnam
the Vietnam Academy of Science and Technology (VAST) under Grants
No.~NVCC.05.12/22-23
and
No.~DL0000.02/24-25.
 
% Policy from October 20, 2022
These acknowledgements are not to be interpreted as an endorsement of any statement made
by any of our institutes, funding agencies, governments, or their representatives.
 
We thank the SuperKEKB team for delivering high-luminosity collisions;
the KEK cryogenics group for the efficient operation of the detector solenoid magnet and IBBelle on site;
the KEK Computer Research Center for on-site computing support; the NII for SINET6 network support;
and the raw-data centers hosted by BNL, DESY, GridKa, IN2P3, INFN, 
PNNL/EMSL, 
and the University of Victoria.
We warmly thank Fu-Sheng Yu and Zhen-Hua Zhang for valuable and helpful discussions.

% Bibliography

%% [A] Recommended: using JHEP.bst file
\bibliographystyle{JHEP}
\bibliography{references.bib}

%% or
%% [B] Manual formatting (see below)
%% (i) We suggest to always provide author, title and journal data or doi:
%% in short all the informations that clearly identify a document.
%% (ii) please avoid comments such as "For a review'', "For some examples",
%% "and references therein" or move them in the text. In general, please leave only references in the bibliography and move all
%% accessory text in footnotes.
%% (iii) Also, please have only one work for each \bibitem.

\end{document}